\documentclass[prd,superscriptaddress,nofootinbib,preprint]{revtex4-1}
\usepackage{bm, 
  graphicx,
  hyperref,
  color,
  braket,
  array,
  slashed,
  mathrsfs,
  subfig,
epsfig
}

\newcommand{\eqref}[1]{(\ref{#1})}

\newcommand{\be}{\begin{eqnarray}}
\newcommand{\ee}{\end{eqnarray}}
\newcommand{\ba}{\begin{align}}
\newcommand{\ea}{\end{align}}

\newcommand{\lsim}{\lesssim}
\newcommand{\gsim}{\gtrsim}

\begin{document}
\begin{flushright}
MAN/HEP/2014/05\\
 IPPP/14/43, DCPT/14/86\\
\today
\end{flushright}
\title{TeV Scale Left-Right Symmetry and Large Mixing Effects in  Neutrinoless Double Beta Decay}

\author{P. S. Bhupal Dev}
\affiliation{Consortium for Fundamental Physics, School of Physics and Astronomy, University of Manchester, Manchester M13 9PL, United Kingdom}

\author{Srubabati Goswami}
\affiliation{Physical Research Laboratory, Navrangpura, Ahmedabad  380009, India}

\author{Manimala Mitra}
\affiliation{Institute for Particle Physics Phenomenology, 
Department of Physics, Durham University, Durham DH1 3LE, United Kingdom}
\begin{abstract}
We analyze  various contributions to neutrinoless double beta decay ($0\nu\beta\beta$) 
in a TeV-scale Left-Right Symmetric Model (LRSM) for type-I seesaw dominance.  We find that 
the momentum-dependent effects due to  $W_L-W_R$ 
exchange  ($\lambda$-diagram) and $W_L-W_R$ mixing ($\eta$-diagram) could give dominant 
contributions to the $0\nu\beta\beta$ amplitude in a wide range of the LRSM parameter space. In particular, for a relatively large $W_L-W_R$ mixing, the $\eta$-contribution by itself could saturate the current experimental limit on the $0\nu\beta\beta$ half-life, thereby providing  stringent constraints on the relevant LRSM parameters, complementary to the indirect constraints derived from lepton flavor violating observables.    
In a simplified scenario parametrized by a 
single light-heavy neutrino mixing, the inclusion of the $\lambda$ and  $\eta$ contributions leads to significantly improved $0\nu\beta\beta$ constraints on the light-heavy neutrino mixing as well as on the $W_L-W_R$ mixing parameters. We also present a concrete TeV-scale LRSM setup, where the mixing effects are manifestly 
enhanced, and discuss the interplay between $0\nu\beta\beta$, lepton flavor violation and electric dipole moment constraints.  
\end{abstract}

\maketitle
\section{Introduction} \label{sec:I}
The process of neutrinoless double beta decay ($0\nu\beta\beta$): $^A_Z X \to ^A_{Z+2}Y+2e^-$, if observed, would be an unambiguous evidence of lepton number violation (LNV), thus confirming the Majorana nature of neutrinos~\cite{Schechter:1981bd}. In addition, it can possibly shed light on some of the yet unresolved issues in neutrino physics, such as the absolute neutrino mass scale, the mass hierarchy, and the origin of tiny neutrino masses; for reviews, see e.g.~\cite{review1}. The current 
lower limit on the $0\nu\beta\beta$ half-life in various nuclei, most notably in $^{136}$Xe by KamLAND-Zen~\cite{kamland} and in $^{76}$Ge by GERDA-I~\cite{gerda}, can be saturated by the canonical light neutrino contribution~\cite{canon} {\em only} in the quasi-degenerate region with masses $m_1 \simeq m_2 \simeq m_3 \gsim 0.1$ eV; for a recent analysis with the updated nuclear matrix elements (NMEs), see~\cite{DGMR}. This is,  however, in conflict with the most stringent upper limit on the sum of light neutrino masses, $\sum_i m_i < 0.23$ eV at 95\% confidence level (CL), obtained from Planck data~\cite{planck}. Hence, any evidence of a positive signal in the upcoming $0\nu\beta\beta$ experiments~\cite{future},\footnote{The only claimed observation of $0\nu\beta\beta$ so far~\cite{klapdor}  is in direct conflict with the GERDA-I results~\cite{gerda} and is also incompatible with the KamLAND-Zen results~\cite{kamland} for most of the NME calculations~\cite{DGMR}.} could indicate a new physics contribution to this rare LNV process.  

One of the simplest paradigms for understanding the observed smallness of neutrino 
masses is the type-I seesaw mechanism~\cite{type1, type1b, mixed}, where SM-singlet heavy 
Majorana neutrinos are introduced. 
If sufficiently light ($\lsim$ 10 TeV), they can give a significant contribution to 
$0\nu\beta\beta$ through their mixing with the active neutrinos~\cite{ibarra, MSV, pascoli, Pascoli:2013fiz}. However, 
in the simplest scenario, which we will call the `SM seesaw', 
there are no guidelines either for the mass scale of the heavy neutrinos or the light-heavy neutrino mixing, and these quantities have to be set `by hand'
in an adhoc manner. 
An attractive theoretical framework, which provides a natural setting for the 
seesaw mechanism, is the Left-Right Symmetric Model (LRSM), 
based on the gauge group 
$SU(2)_L\times SU(2)_R\times U(1)_{B-L}$~\cite{LR}. 
In this model, the SM-singlet 
neutrino fields  are inducted as the necessary parity gauge partners, i.e. the right-handed (RH) counterparts, of the corresponding left-handed (LH) neutrino fields, whereas the seesaw scale is intimately connected to the 
$SU(2)_R\times U(1)_{B-L}$-breaking scale.  The LRSM can naturally explain the small neutrino masses through either type-I seesaw via the RH neutrinos~\cite{type1, type1b} or type-II seesaw via $SU(2)$-triplet scalars~\cite{type2, type2b} or both~\cite{mixed}.

In LRSM, there are several new contributions to $0\nu\beta\beta$, involving RH neutrinos and RH gauge bosons~\cite{type1b, type2b}, Higgs triplets~\cite{MV}, 
as well as mixed LH-RH contributions~\cite{hirsch} (for some recent studies, 
see e.g.~\cite{chakra, barry, huang}). 
A general analysis of $0\nu\beta\beta$ in the LRSM including all
the diagrams is rather complicated. For a simplified case with type-II seesaw 
dominance, the light-heavy neutrino mixing  is negligible, and the   
light neutrino mass matrix is directly proportional to the heavy neutrino 
mass matrix, with the constant of proportionality given by the ratio of 
the LH and RH triplet-scalar vacuum expectation values (VEVs). In this case, 
the dominant new contribution to the $0\nu\beta\beta$ process comes from the diagram 
with purely RH currents involving the heavy gauge boson $W_R$ and the heavy neutrinos~\cite{Tello:2010am}. 
The current limit on the half-life of $0\nu \beta \beta$  can be saturated by this new contribution alone, which however puts a {\it lower} limit on the lightest neutrino mass~\cite{DGMR}, {as long as the heavy neutrino masses in the LRSM are well above the typical momentum exchange scale $\sim 100$ MeV}. 
In addition, $0\nu\beta\beta$ also provides constraints on the RH gauge boson and heavy neutrino masses~\cite{Nemevsek:2011aa, DGMR, das}, which are complementary to the limits obtained from  direct searches at the LHC~\cite{LHC-RR}, from low-energy lepton flavor violating (LFV) observables~\cite{das, barry}, as well as from hadronic flavor and $C\!P$ violating effects~\cite{fcnc, Bertolini:2014sua}. 

For the case of type-I seesaw dominance, 
the light neutrino mass matrix is dominantly generated by the Dirac mass matrix $M_D$ and the RH neutrino mass matrix $M_R$ through the usual seesaw formula: 
\be
M_\nu \ \simeq \ -M_D M_R^{-1} M_D^{\sf T}  \;.
\label{eq1}
\ee
 In this case, 
there are additional contributions to $0\nu \beta \beta$, that involve the light-heavy neutrino mixing 
parameter $\theta \simeq M_DM_R^{-1}$. In the canonical seesaw, the observed smallness of light neutrino masses puts severe constraints on this mixing parameter. For instance, for a TeV-scale RH neutrino mass, the mixing angle is required to be $\lsim {\cal O}(10^{-6})$, in order to reproduce the 
sub-eV scale active neutrino masses. 
However, in the presence of cancellations in the matrix structure on the RHS of Eq.~\ref{eq1}, this constraint can be significantly relaxed, and the light neutrino oscillation data can be satisfied even with a larger value of $\theta$~\cite{cancel1, ibarra, MSV}. 
This has potentially huge implications for the experimental tests of the SM seesaw at colliders~\cite{theory-LL} as well as in other low-energy experiments (for reviews, see e.g.~\cite{review2}). 

A large light-heavy neutrino mixing is also possible in LRSM~\cite{cancel, DLM}. 
In this case,  there are further additional contributions to $0\nu \beta \beta$, involving RH neutrino and/or RH gauge boson exchange, which could be significant~\cite{MSV, barry, chakra, Vergados:2002pv}.  
In particular, the mixed diagrams  involving LH-RH currents and with final state electrons of opposite helicity, known as the $\lambda$ and $\eta$ diagrams, could be important~\cite{barry, huang, parida}. Specifically, the $\eta$-contribution, which depends on the light-heavy neutrino mixing parameter $\theta$ as well as the LH-RH gauge boson mixing parameter $\xi$, can be sizable~\cite{DLM} as the NMEs for
the $\eta$-diagram are roughly two orders of magnitude larger than those for the 
$\lambda$-diagram~\cite{Pantis:1996py, Suhonen:1998ck}.   
As we will show in this paper, for a relatively larger 
value of $\xi$ close to its current experimental upper bound, the $\eta$-diagram could give the dominant contribution to $0\nu\beta\beta$. 
Note that experimentally, the different mechanisms for $0\nu\beta\beta$ in LRSM could be potentially discriminated by measuring the electron angular and energy distributions in the upcoming SuperNEMO experiment~\cite{sune}. Moreover, 
a  large light-heavy neutrino mixing in LRSM also gives an 
additional contribution to the like-sign dilepton 
signal~\cite{KS} at the LHC: $pp\to W_R \to N\ell^\pm \to \ell^\pm \ell^\pm W_L^\mp$~\cite{chendev}, and also to the inverse $0\nu\beta\beta$ process $e^-e^-\to W_L^- W_R^-$ at the ILC~\cite{Barry:2012ga}, thus enhancing the prospects of directly probing the seesaw mechanism at colliders.  

In this paper, we carefully analyze all relevant contributions to 
$0\nu\beta\beta$ in the LRSM for type-I seesaw dominance.  
Specifically, we emphasize the importance of the $\lambda$ and $\eta$ contributions 
mentioned above and explicitly demonstrate that, for relatively 
large $\xi$ values, the $\eta$-contribution by itself can saturate the 
current lower limit on the $0\nu\beta\beta$ half-life. Working within a simplified scenario, 
parametrized by a single RH neutrino mass scale $M_R$ and a single light-heavy neutrino mixing angle $\theta$, we show that the constraints from $0\nu\beta\beta$ process including the $\lambda$ and $\eta$ contributions leads to an improved upper bound on the light-heavy neutrino mixing parameter in certain ranges of the parameter space. 
 Comparing this model-independent bound with the complementary constraints from LFV 
observables, we find that, for a given value of $\xi$, the $0\nu\beta\beta$ 
constraint on $\theta$ could be the most stringent one. Using the current lower limits on the 
$0\nu\beta\beta$ half-life from GERDA and KamLand-Zen, we also derive an 
upper limit on the mixing parameter $\xi$, which is much stronger than the 
existing limit~\cite{pdg} in a 
wide range of the LRSM parameter space. Finally, we consider a concrete low-scale 
type-I seesaw scenario with large light-heavy neutrino mixing, and show the importance of the mixed contributions on $0\nu\beta\beta$. We also study the interplay of $0\nu\beta\beta$ with the LFV and electric dipole moment predictions within this framework.

The paper is organized as follows: In Section~\ref{rev}, we review the basic features of the minimal LRSM. The different contributions to the $0\nu \beta \beta$ amplitude in the type-I seesaw dominance are discussed in Section~\ref{diffcont}. In Section~\ref{gen}, we consider a simplified case with a single light-heavy neutrino mixing parameter $\theta$ and  derive improved upper limits on $\theta$ from $0\nu \beta \beta$ constraints. In addition, we also derive an improved upper limit on the LH-RH gauge boson mixing parameter $\xi$ as a function of $\theta$. In Section~\ref{threerh}, we discuss a general case with three RH neutrino flavors.  
 In Section~\ref{smallmix}, we first consider  the case where the light-heavy neutrino mixing is small, and the dominant contribution to $0\nu\beta\beta$ comes from the purely RH sector. In Section~\ref{largemix}, we present a specific TeV-scale seesaw model with large mixing and explicitly show the importance of the $\eta$ and $\lambda$ contributions. Our conclusions are given in Section~\ref{conclu}.

\section{Review of the Minimal Left-Right Symmetric Model} 
\label{rev}
For completeness and to set our notations, we review the basic features of the minimal LRSM, based on the gauge group $SU(3)_c\times SU(2)_L\times SU(2)_R\times U(1)_{B-L}\equiv G_{3221}$~\cite{LR}. The quarks and leptons are assigned to the following irreducible representations of the gauge group $G_{3221}$: 
\be
& Q_{L,i} \ = \ \left(\begin{array}{c}u_L\\d_L \end{array}\right)_i : \: \left({ \bf 3}, {\bf 2}, {\bf 1}, \frac{1}{3}\right), \qquad \qquad  
Q_{R,i} \ = \ \left(\begin{array}{c}u_R\\d_R \end{array}\right)_i : \: \left({ \bf 3}, {\bf 1}, {\bf 2}, \frac{1}{3}\right), \nonumber \\
& \psi_{L,i} \ = \  \left(\begin{array}{c}\nu_L \\ e_L \end{array}\right)_i : \: \left({ \bf 1}, {\bf 2}, {\bf 1}, -1 \right), \qquad \qquad 
\psi_{R,i} \ = \ \left(\begin{array}{c} N_R \\ e_R \end{array}\right)_i : \: \left({ \bf 1}, {\bf 1}, {\bf 2}, -1 \right),
\ee
where $i=1,2,3$ is the family index, and the subscripts $L,R$ are associated with the left and right chiral projection operators $P_{L,R} = (1\mp \gamma_5)/2$. The electric charge is given by $Q=I_{3L}+I_{3R}+(B-L)/2$, where $I_{3L}$ and $I_{3R}$ are the third components of isospin under $SU(2)_L$ and $SU(2)_R$ respectively.  
For the scalar sector, we must choose L-R symmetric Higgs multiplets. The first choice is a bi-doublet under $SU(2)_L\times SU(2)_R$:
\be
\Phi = \left(\begin{array}{cc}\phi^0_1 & \phi^+_2\\\phi^-_1 & \phi^0_2\end{array}\right) : ({\bf 1}, {\bf 2}, {\bf 2}, 0),
\ee
which couples to the fermion bilinears $\bar{Q}_LQ_R$ and $\bar{\psi}_L\psi_R$, and gives masses to quarks and leptons after spontaneous symmetry breaking by its VEV: $\langle\Phi\rangle={\rm diag}(\kappa_1, \kappa_2)/\sqrt{2}$. However, since $\Phi$ is neutral under $B-L$, its VEV  cannot break the $U(1)_{B-L}$-symmetry. In the minimal LRSM, the L-R symmetry is broken by an additional pair of $SU(2)$ triplets:  
\be
 \Delta_L\equiv\left(\begin{array}{cc}\Delta^+_L/\sqrt{2} & \Delta^{++}_L\\\Delta^0_L & -\Delta^+_L/\sqrt{2}\end{array}\right) : ({\bf 1}, {\bf 3}, {\bf 1}, 2), \qquad 
\Delta_R\equiv\left(\begin{array}{cc}\Delta^+_R/\sqrt{2} & \Delta^{++}_R\\\Delta^0_R & -\Delta^+_R/\sqrt{2}\end{array}\right) : ({\bf 1}, {\bf 1}, {\bf 3}, 2),  
\end{eqnarray}
which also give the Majorana mass terms for heavy neutrinos. 

The gauge symmetry $SU(2)_R\times U(1)_{B-L}$ is broken down to the group $U(1)_Y$ of the SM by the VEV of the neutral component of $\Delta_R$: $\langle \Delta^0_R\rangle = v_R/\sqrt{2}$. Since this gives masses to the RH gauge bosons $W_R$ and $Z'$, the current experimental limits~\cite{pdg} suggest $v_R\gsim 6$ TeV. There is also an LH counterpart $\langle \Delta^0_L \rangle = v_L/\sqrt{2}$, which however is required to be small: $v_L\lsim 5$ GeV due to the $\rho$-parameter constraints~\cite{pdg}. 
Finally, the VEV of the $\Phi$ field breaks the SM gauge group $SU(2)_L\times U(1)_Y$ to $U(1)_Q$, and hence, is expected to be at the electroweak scale. Thus we have the following hierarchy of VEVs: 
\be
v_L\ll \kappa_{1,2} \ll v_R \; .
\label{eq:hie}
\ee 
Making use of the gauge symmetry, we can eliminate some of the complex phases in the scalar sector, and treat $\kappa_1$ and $v_R$ as real, while $\kappa_2$ and $v_L$ are, in general, complex parameters.

The Yukawa Lagrangian in the lepton sector is given by 
\begin{eqnarray}
-{\cal L}_Y &=& h_{ij}\bar{\psi}_{L,i}\Phi \psi_{R, j} 
+ \tilde{h}_{ij}\bar{\psi}_{L, i}\tilde{\Phi} \psi_{R, j}
+ f_{L, ij} \psi_{L,i}^{\sf T} C i\tau_2 \Delta_L \psi_{L,j} 
+ f_{R, ij} \psi_{R,i}^{\sf T} C i\tau_2 \Delta_R \psi_{R,j} 
+{\rm H.c.},
\label{eq:yuk}
\end{eqnarray}
where the family indices $i,j$ are summed over, $C=i\gamma_2\gamma_0$ is the charge conjugation operator, and $\tilde{\Phi}=\tau_2\Phi^*\tau_2$, with $\tau_2$ being the second Pauli matrix, and $\gamma_\mu$ the Dirac matrices. After symmetry breaking, 
Eq.~\eqref{eq:yuk} leads to the following $6\times 6$ neutrino mass matrix:
\be
{\cal M}_\nu = \left(\begin{array}{ccc}
M_L & M_D \\
M_D^{\sf T} & M_R
\end{array}\right), 
\label{eq:big}
\ee
where the $3\times 3$ Dirac and Majorana mass matrices are given by 
\be 
M_D = \frac{1}{\sqrt 2}\left(\kappa_1 h + \kappa_2 \tilde{h} \right), \quad 
M_L = \sqrt 2 v_L f_L, \quad
M_R = \sqrt 2 v_R f_R \; .
\ee
For the hierarchy of VEVs given by Eq.~\eqref{eq:hie}, the $3\times 3$ light neutrino 
mass matrix becomes 
\be 
M_\nu \ \simeq \ M_L - M_D M_R^{-1} M_D^{\sf T} 
 \ = \ \sqrt 2 v_L f_L - \frac{\kappa^2}{\sqrt 2 v_R} h_D f_R^{-1} h_D^{\sf T} \; , 
\label{eq:mnu}
\ee
where $h_D \equiv (\kappa_1 h+\kappa_2 \tilde{h})/(\sqrt 2 \kappa)$ and $\kappa \equiv (|\kappa_1|^2+|\kappa_2|^2)^{1/2}$. 
 Note that, 
 in the type-I dominance, we set $v_L=0$ \footnote{The minimization of the general LRSM potential can allow this option~\cite{goran_vLzero}.}.
Consequently the first term on the RHS of Eq.~\eqref{eq:mnu} can be dropped
and one recovers the usual type-I seesaw formula given by Eq.~\eqref{eq1}.  

The full neutrino mass matrix in Eq.~\eqref{eq:big} can be diagonalized by a $6\times 6$ unitary matrix, as follows: 
\be
{\cal V}^{\sf T}{\cal M}_\nu {\cal V} \ = \ \left(\begin{array}{cc} \widehat{M}_\nu & {\bf 0} \\ {\bf 0} & \widehat{M}_R \end{array}\right) ,
\ee
where $\widehat{M}_\nu = {\rm diag}(m_1,m_2,m_3)$ and $\widehat{M}_R = {\rm diag}(M_1,M_2,M_3)$. The unitary matrix ${\cal V}$ has an exact representation in terms of an arbitrary $3\times 3$ matrix $\zeta$~\cite{Korner:1992zk,Grimus:2000vj}: 
\be
{\cal V} \ = \ \left(\begin{array}{cc}
({\bf 1}+\zeta^*\zeta^{\sf T})^{-1/2} & \zeta^*({\bf 1}+\zeta^{\sf T}\zeta^*)^{-1/2} \\
-\zeta^{\sf T}({\bf 1}+\zeta^*\zeta^{\sf T})^{-1/2} & ({\bf 1}+\zeta^{\sf T}\zeta^*)^{-1/2}
\end{array}  \right)
\left(\begin{array}{cc}
U_\nu & {\bf 0} \\
{\bf 0} & V_R
\end{array}\right) \ \equiv \ \left(\begin{array}{cc}
U & S \\
T & V 
\end{array}\right) \; ,
\label{V}
\ee
where $\zeta^*= M_DM_R^{-1}$ to leading order in a converging Taylor series expansion, and $U_\nu$, $V_R$ are the $3\times 3$ unitary matrices diagonalizing the light and heavy neutrino mass 
matrices $M_\nu$ and $M_R$: 
\be 
U^{\sf T}_{\nu} M_\nu U_{\nu} \ = \ \widehat{M}_\nu \; , \qquad \qquad 
V_R^{\sf T} M_R V_R \ = \ \widehat{M}_R \; .
\label{diag}
\ee
The order parameter of the light-heavy neutrino mixing is given by the norm $\|\zeta\| = \sqrt{{\rm Tr}(\zeta^\dag \zeta)} \equiv \theta$, which also measures the non-unitarity of the light neutrino mixing matrix $U$. From Eq.~(\ref{V}), we obtain $S= M_D M^{-1}_R V_R$ and  $T=- (M_D M^{-1}_R)^{\dagger} U_{\nu}$ up to ${\cal O}(\|\zeta\|^2)$.

In the gauge sector, assuming manifest L-R symmetry so that $g_L=g_R\equiv g$ for the $SU(2)$ gauge couplings, the charged gauge boson mass matrix is given by 
\be
{\cal M}_W = \frac{g^2}{4}\left(\begin{array}{cc}
\kappa_1^2+\kappa_2^2+2v_L^2 & 2\kappa_1\kappa_2 \\
2\kappa_1\kappa_2 & \kappa_1^2+\kappa_2^2+2v_R^2
\end{array}\right), 
\ee
with the mass eigenstates 
\be
\left(\begin{array}{c}
W_1 \\
W_2 
\end{array}\right) = \left(\begin{array}{cc}
\cos \xi & \sin \xi \\
-\sin \xi & \cos \xi 
\end{array}\right) 
\left(\begin{array}{c}
W_L \\
W_R 
\end{array}\right),
\ee
where the $W_L-W_R$ mixing parameter is defined by 
\be
\tan{2\xi} = \frac{2\kappa_1 \kappa_2}{v_R^2-v_L^2} . 
\label{kappa}
\ee
For $\xi\ll 1$, the gauge boson masses are given by 
\be
M_{W_1} \simeq M_{W_L} \simeq \frac{g}{2}\kappa, \qquad M_{W_2} \simeq M_{W_R} \simeq \frac{g}{\sqrt 2}v_R .
\ee
In what follows, we will assume $\kappa_2\ll \kappa_1$, although this is strictly not a phenomenological requirement. In this limit, Eq.~\eqref{kappa} can be written as 
\be
\xi \simeq \frac{\kappa_1\kappa_2}{v_R^2} \simeq \frac{2\kappa_2}{\kappa_1}\left(\frac{M_{W_L}}{M_{W_R}}\right)^2 ,
\label{xiup}
\ee 
and hence, $\xi$ is bounded above by $(M_{W_L}/M_{W_R})^2$. Experimentally, the electroweak precision data (EWPD) puts an upper bound on $\xi< 0.013$~\cite{Langacker:1989xa, Czakon:1999ga}, which tightens to $\xi<0.0025$~\cite{Langacker:1989xa} if the $C\!P$-violating phases in the mixing matrix for RH quarks are small.\footnote{For $M_R\lsim 10$ MeV, the upper limit derived from supernova data is even more stringent: $\xi< 10^{-5}$~\cite{Raffelt:1987yt}.}  As far as the RH gauge bosons are concerned, flavor and $C\!P$ violating processes in $K$ and $B$ meson mixing provide an absolute lower bound on $M_{W_R}\gsim 2.9$ TeV~\cite{Bertolini:2014sua}. Complementary bounds of similar magnitude were also obtained from direct searches for the same-sign dilepton signal~\cite{KS} at the LHC~\cite{LHC-RR}. Using these limits, we obtain from Eq.~(\ref{xiup}), 
$\xi \lsim 7.7\times 10^{-4}$.  

\section{$0\nu\beta\beta$ in LRSM \label{diffcont}}
In this section, we briefly discuss the relevant contributions to the $0\nu \beta \beta$ process in a 
TeV-scale LRSM with type-I seesaw dominance (for a detailed discussion, see e.g.~\cite{barry}).
\begin{itemize}
\item [(a)] Light neutrino contribution: This is a purely LH contribution mediated by light Majorana neutrinos, as shown in Figure~\ref{fig1}(a). The corresponding amplitude is given by  
\be 
\mathcal{A}_{\nu} \ \simeq \  G^2_F \sum_i U^2_{ei} \frac{m_i}{p^2} \ \equiv \ 
G^2_F \frac{m_{ee}}{p^2} \; , 
\label{Anu}
\ee 
where $|p|\sim 100$ MeV is the typical momentum transfer at the leptonic vertex, $G_F$ is the Fermi coupling constant and $m_{ee}\equiv \sum_i U^2_{ei} m_i$ is the effective neutrino mass. The light neutrino contribution by itself can  saturate the current experimental limit {\it only} in the quasi-degenerate region with $m_i\simeq 0.1$ eV, which is almost ruled out by the recent Planck data on the sum of light neutrino masses; see~\cite{DGMR} for a recent discussion including the updated NME uncertainties.

\item [(b)] Heavy neutrino contribution: This is the RH counterpart of the purely LH contribution discussed above, and is shown in Figure~\ref{fig1}(b). In this case, assuming that the heavy neutrino mass scale is larger than the momentum exchange scale, i.e.~$M^2_i\gg |p|^2$, we obtain 
\be 
\mathcal{A}^R_{N_R} \ \simeq \ G^2_F \left(\frac{M_{W_L}}{M_{W_R}}\right)^4 \sum_i  \frac{{V_{ei}^*}^2}{M_i} \; .
\label{RR}
\ee
Note that this contribution is independent of the light-heavy neutrino mixing, and hence, the only dominant contribution in the small mixing limit $\theta\to 0$, which could saturate the current experimental limit for smaller values of $M_i$.

\item [(c)] Light-heavy neutrino mixing contribution: A large light-heavy neutrino mixing can induce an additional contribution due to heavy neutrino exchange with purely LH currents, as shown in Fig.~\ref{fig1}(c). 
The amplitude of this process is given by 
\be \mathcal{A}^L_{N_R} \ \simeq \ G^2_F \sum_i \frac{S^2_{ei}}{M_i} \; .
\label{lnr} 
\ee 
Note that this contribution is present even in the minimal SM seesaw scenario without the L-R symmetry, and can be large in presence of cancellations in the light neutrino mass matrix~\cite{MSV}. In LRSM, in addition to the process (c), there is an analogous contribution due to light neutrino exchange with purely RH currents; however, this is highly suppressed by a factor of 
$(M_{W_L}/M_{W_R})^2(M_im_i/|p|^2)\lsim 10^{-9}$, as compared to the process (c), and therefore, is not shown in Figure~\ref{fig1}.  
\begin{figure}[t!]
\centering
\includegraphics[width=5.2cm]{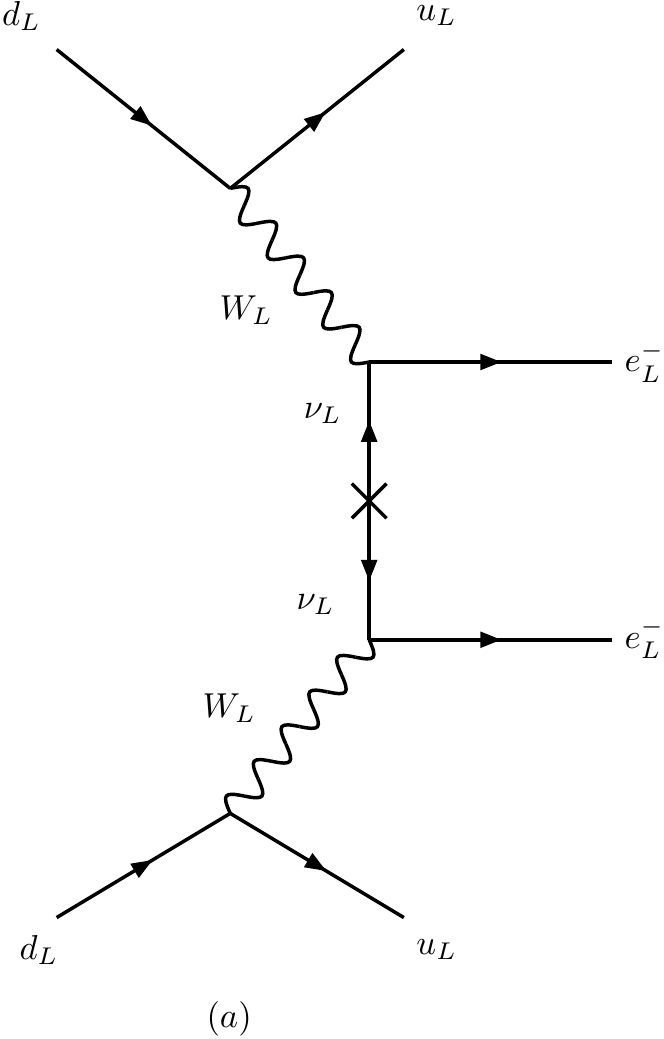}
\includegraphics[width=5.2cm]{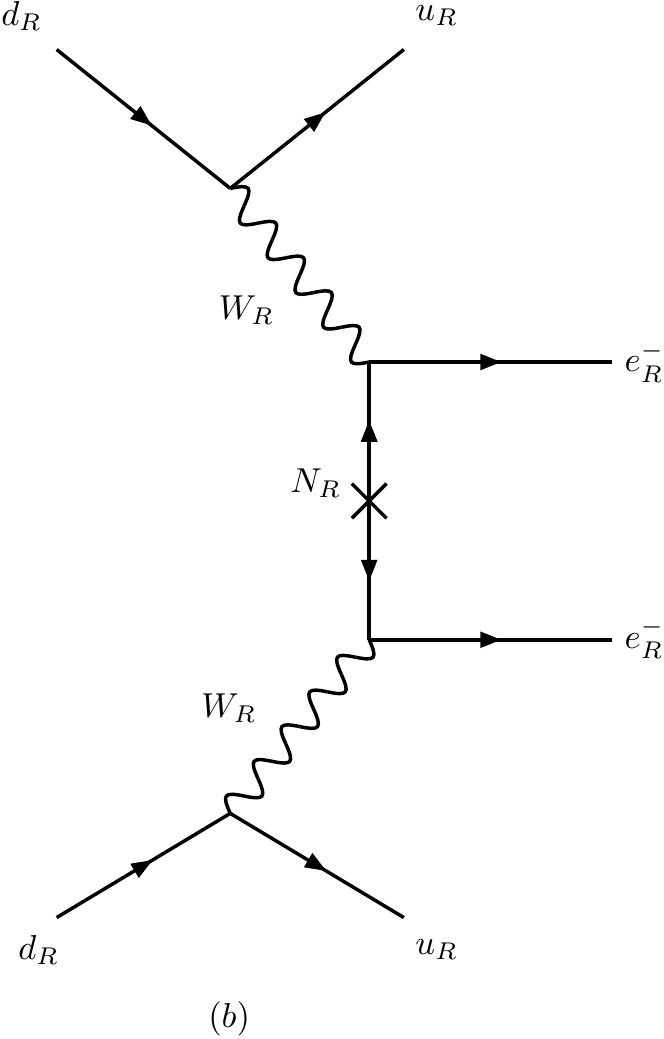}
\includegraphics[width=5.2cm]{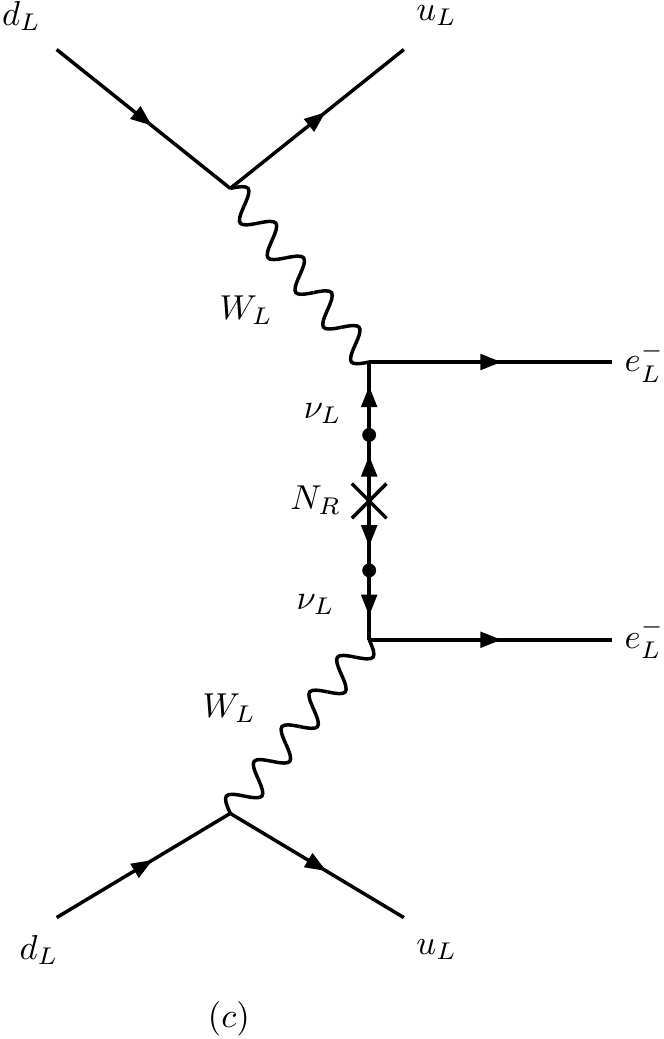} \\ [0.4cm]
\includegraphics[width=5.2cm]{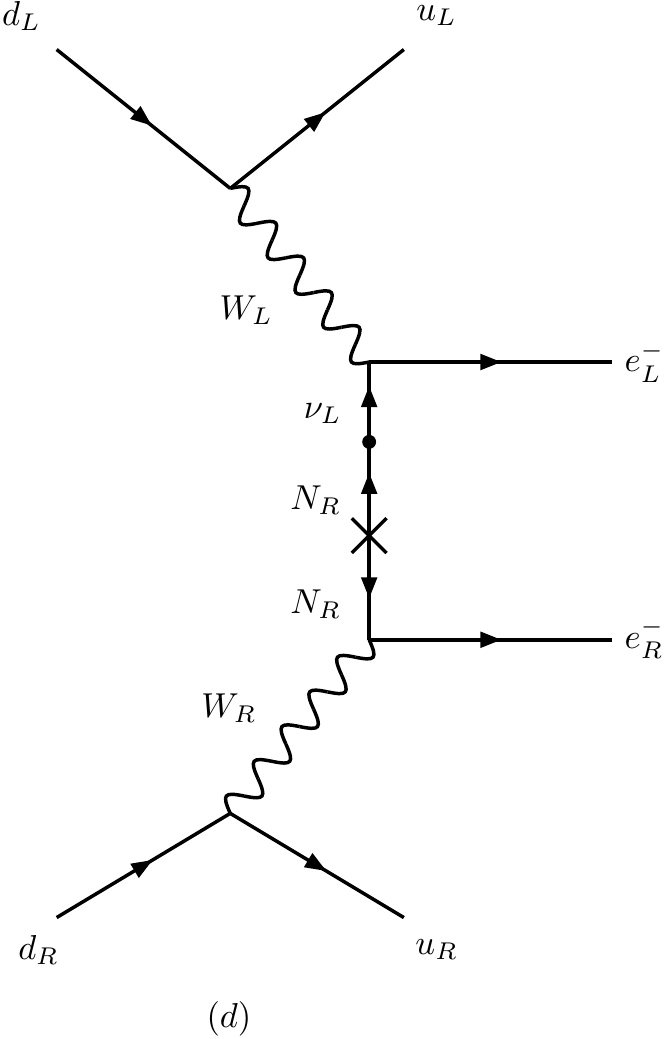}
\includegraphics[width=5.2cm]{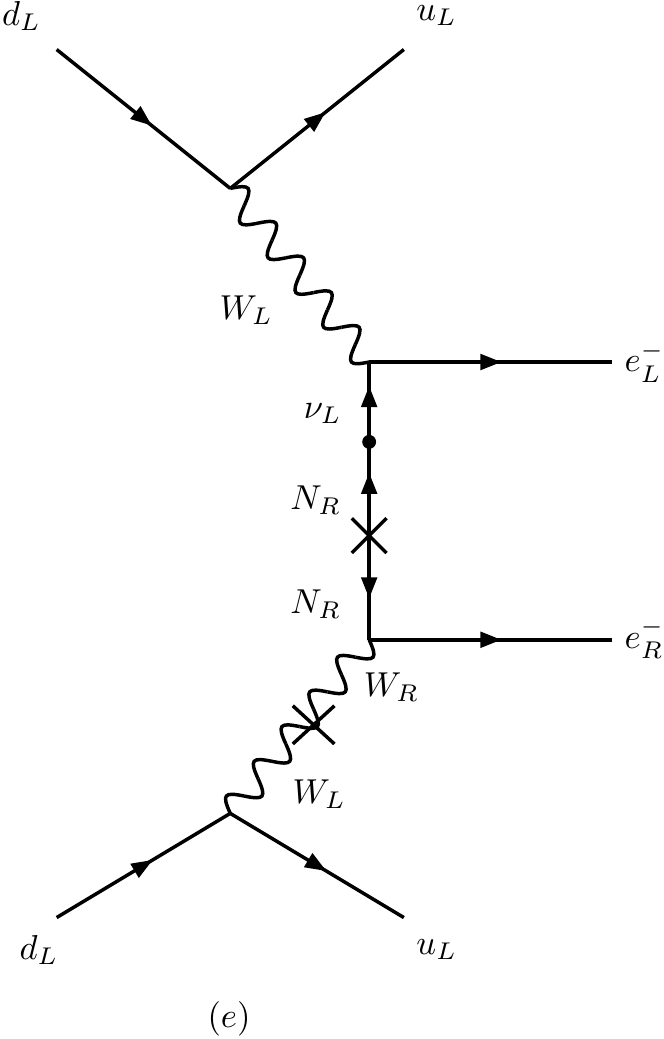}
\caption{Feynman diagrams for the relevant contributions to $0\nu\beta\beta$ in the LRSM with type-I seesaw dominance: (a) ${\cal A}_\nu$, (b) ${\cal A}_{N_R}^R$, (c) ${\cal A}_{N_R}^L$, (d) ${\cal A}_{\lambda}$, and (e) ${\cal A}_\eta$. The $\times$ symbol on the fermion propagator denotes the Majorana mass insertion, while the~\textbullet~symbol denotes light-heavy neutrino mixing. The $\times$ symbol on the gauge boson propagator in diagram (e) denotes the $W_L-W_R$ mixing. }
\label{fig1}
\end{figure}

All the processes (a)-(c) mentioned above  involve final state electrons with the same helicities, i.e.~$e^-_Le^-_L$ or $e^-_Re^-_R$. In the LRSM, there are two important additional processes induced by light-heavy mixing effects, which involve final state electrons with opposite helicities $e^-_Le^-_R$, as discussed below.   

\item [(d)] $\lambda$ contribution:  This process is mediated by $W_L- W_R$ exchange, as shown in Fig.~\ref{fig1}(d). The amplitude is given by  
\be
\mathcal{A}_{\lambda} \ \simeq \ G^2_F \left(\frac{M_{W_L}}{M_{W_R}}\right)^2 \sum_i U_{ei} T^*_{ei} \frac{1}{p} \; .
\label{lam}
\ee

\item [(e)] $\eta$ contribution: This process depends on the $W_L$--$W_R$ mixing parameter $\xi$, as shown in Figure~\ref{fig1}(e). The amplitude for this process is given by 
\be
\mathcal{A}_{\eta} \ \simeq \ G^2_F \tan \xi \,{\sum_i U_{ei} T^*_{ei}}\frac{1}{p} \; .
\label{eta}
\ee
\end{itemize}

From Eqs.~(\ref{lam}) and (\ref{eta}), we see that both $\lambda$ and $\eta$ contributions 
depend on the same combination of the mixing matrices  $U$ and $T$. Hence, if the $\lambda$ contribution is large in some region of the model parameter space, the $\eta$ contribution should also be large for reasonably large values of 
the gauge boson mixing parameter $\xi$, and therefore, cannot be neglected in general. Moreover, the ratio of the NMEs corresponding to 
$\eta$ and $\lambda$ diagrams is of $\mathcal{O}(10^{2})$~\cite{Pantis:1996py}. Hence, even for a moderately lower  value of $\xi$, the $\eta$ contribution can be comparable to or larger than the $\lambda$ contribution. As we will show below, for larger $\xi$ values close to its current experimental limit of $\sim 10^{-3}$, the $\eta$ contribution is indeed the dominant one in a wide range of LRSM parameter space. 

Apart from the diagrams shown in Figure~\ref{fig1}, there could be additional contributions to $0\nu\beta\beta$ in LRSM due to the Higgs triplets. However, in the type-I seesaw dominance, we assume the left-triplet VEV to be negligible, which implies the contribution from the diagram mediated by the $SU(2)_L$ triplet can be ignored. Moreover, the $SU(2)_R$ triplet is required to be heavy to suppress the tree-level LFV process $\mu^-\to e^-e^+e^-$~\cite{Tello:2010am}. We assume this to be the case, and hence, do not consider the triplet contributions in our subsequent analysis. 

Combining all the above processes (a)--(e), one obtains the following expression for the half-life of the $0\nu\beta\beta$ process for a given nuclear isotope: 
\be
\frac{1}{T_{1/2}^{0\nu}} \ = \ G_{01}^{0\nu} \Bigg(\Big|{\cal M}_\nu^{0\nu}\eta_\nu+{\cal M}_N^{0\nu}\eta_{N_R}^L\Big|^2 + \Big|{\cal M}_N^{0\nu}\eta_{N_R}^R\Big|^2 + \Big|{\cal M}_\lambda^{0\nu}\eta_\lambda + {\cal M}_\eta^{0\nu}\eta_\eta\Big|^2\Bigg)  \; ,
\label{half}
\ee   
where $G_{01}^{0\nu}$ is the phase space factor and ${\cal M}^{0\nu}_X$ are the relevant NMEs, whose numerical values for $^{76}$Ge and $^{136}$Xe nuclei are given in Table~\ref{tab1}. The $\eta$'s are the dimensionless particle physics parameters obtained from the Feynman amplitudes given in Eqs.~(\ref{Anu})-(\ref{eta}), as follows: 
\be
\eta_\nu & = & \frac{1}{m_e}\sum_i U_{ei}^2 m_i \; , \label{etanu}\\
\eta^R_{N_R} & = & m_p \left(\frac{M_{W_L}}{M_{W_R}}\right)^4 \sum_i  \frac{{V_{ei}^*}^2}{M_i} \; , \label{etaRR} \\
\eta^L_{N_R}  &=&  m_p \sum_i \frac{S^2_{ei}}{M_i} \; , \label{etaRL} \\
\eta_{\lambda} &=&  \left(\frac{M_{W_L}}{M_{W_R}}\right)^2 \sum_i U_{ei} T^*_{ei} \; , \label{etalam} \\
\eta_{\eta} &=&  \tan \xi \sum_i U_{ei} T^*_{ei} \; , \label{etaeta}
\ee
where $m_e$ and $m_p$ are the masses of electron and proton, respectively. In Eq.~(\ref{half}), we have included the  interference effect between diagrams having final state electrons with the same helicity combination. We have neglected the interference terms between the diagrams with different helicity final state pair, which will be suppressed by the electron mass. 

For the phase space factors $G_{01}^{0\nu}$ in Eq.~(\ref{half}), we use the recent calculation of~\cite{Kotila:2012zza} for the axial-vector coupling $g_A=1.25$, whereas for the NMEs in Eq.~(\ref{half}), we use the QRPA calculation of~\cite{Pantis:1996py}. The lower values of the NMEs in Table~\ref{tab1} are obtained for the case without $p$--$n$ pairing, 
whereas the higher values are with $p$--$n$ pairing. Using these values in Eq.~(\ref{half}), and assuming no interference effects, we derive upper limits on the dimensionless parameters given by 
Eqs.~(\ref{etanu})-(\ref{etaeta}) from the current 90\% CL combined lower limits on the half-lives of $^{76}{\rm Ge}$ and $^{136}{\rm Xe}$ obtained from GERDA-I+Heidelberg-Moscow+IGEX~\cite{gerda} and KamLAND-Zen+EXO-200~\cite{kamland}, respectively:   
\be
T_{1/2}^{0\nu}(^{76}{\rm Ge}) \ > \ 3.0\times 10^{25}~{\rm yr}\; , \qquad 
T_{1/2}^{0\nu}(^{136}{\rm Xe}) \ > \ 3.4\times 10^{25}~{\rm yr} \; . \label{GeXe}
\ee
The results are shown in Table~\ref{tab2}, where the range is due to the NME uncertainties. 

Comparing the half-life predictions given by Eq.~(\ref{half}) with the
current experimental limits given by Eq.~(\ref{GeXe}), we derive constraints on the 
LRSM model parameter space in the type-I seesaw dominance, as discussed in the following two sections. 

\begin{table}[t!] 
\begin{center}
\begin{tabular}{c|c|c|c|c|c}\hline\hline
 & $G_{01}^{0\nu}$ & \multicolumn{4}{c}{Nuclear Matrix Elements} \\ \cline{3-6}
Isotope & (yr$^{-1}$) & ${\cal M}_\nu^{0\nu}$ & ${\cal M}_N^{0\nu}$ & ${\cal M}_\lambda^{0\nu}$ & ${\cal M}_\eta^{0\nu}$ \\ \hline
$^{76}$Ge & $5.77\times 10^{-15}$ & 2.58--6.64 & 233--412 & 1.75--3.76 & 235--637 \\
$^{136}$Xe & $3.56\times 10^{-14}$ & 1.57--3.85 & 164--172 & 1.92--2.49 & 370--419 \\ 
\hline\hline
\end{tabular}
\end{center}
\caption{Numerical values of the phase-space factor $G_{01}^{0\nu}$~\cite{Kotila:2012zza} and the NMEs ${\cal M}_X^{0\nu}$~\cite{Pantis:1996py} for different contributions to $0\nu\beta\beta$, as used in our analysis. }
\label{tab1}
\end{table}
\begin{table}[t!] 
\begin{center}
\begin{tabular}{c|c|c|c}\hline\hline
& Dimensionless & \multicolumn{2}{c}{Current Upper Limit} \\ \cline{3-4}
Mechanism & Parameter & $^{76}$Ge & $^{136}$Xe \\
\hline\hline
(a) Light neutrino exchange (LH current) & $|\eta_\nu|$ & $(3.6-9.3)\times 10^{-7}$ & $(2.4-5.8)\times 10^{-7}$ \\
(b) Heavy neutrino exchange (RH current) & $|\eta^R_{N_R}|$ & $(0.58-1.0)\times 10^{-8}$ & $(5.3-5.5)\times 10^{-9}$\\
(c) Heavy neutrino exchange (LH current) & $|\eta^L_{N_R}|$ & $(0.58-1.0)\times 10^{-8}$ & $(5.3-5.5)\times 10^{-9}$ \\
(d) $\lambda$-diagram (LH-RH current) & $|\eta_{\lambda}|$ & $(0.64-1.4)\times 10^{-6}$ & $(3.6-4.6)\times 10^{-7}$ \\
(e) $\eta$-diagram (LH-RH current) & $|\eta_{\eta}|$ & $(0.38-1.0)\times 10^{-8}$ & $(2.2-2.5)\times 10^{-9}$ \\
\hline\hline
\end{tabular}
\end{center}
\caption{The experimental upper limits on the dimensionless particle physics parameters describing various $0\nu\beta\beta$ contributions shown in Figure~\ref{fig1} [cf. Eqs.~(\ref{etanu})-(\ref{etaeta})]. }
\label{tab2}
\end{table}
\section{A General  Analysis for Type-I Dominance \label{gen}}
In this section, we illustrate the relative magnitudes of the different contributions to
$0\nu \beta \beta$ discussed in Section~\ref{diffcont} for a generic LRSM in the type-I seesaw dominance. For this purpose, we consider a simplified generation-independent scenario parametrized by a single
RH neutrino mass scale $M_R$ and a single light-heavy neutrino mixing parameter $\theta$ 
in the electron sector, without specifying the full flavor structures of $M_D$ and $M_R$. In addition, we assume the RH-neutrino mixing matrix to be the same as the LH-neutrino mixing matrix, i.e., $V_R=U_\nu$ in Eq.~(\ref{diag}). For $|\theta|^2\ll 1$, ignoring the non-unitarity of the $3\times 3$ light-neutrino mixing matrix 
$U$ in Eq.~(\ref{V}), we will take $U$ to be the PMNS mixing matrix. Using the best fit values of a recent three-neutrino global analysis~\cite{newosc} for the light neutrino oscillation parameters, and assuming a normal hierarchy of light neutrino masses, we obtain 
\be
U \ = \ V \ = \ \left(\begin{array}{ccc}
0.8221 & 0.5484 & -0.0518-0.1439\: i \\
-0.3879-0.0791\: i & 0.6432-0.0528\: i & 0.6533 \\
0.3992-0.0898\: i & -0.5283-0.0599 \: i & 0.7415
\end{array}\right) \; .
\ee
Here we have taken the Dirac $C\!P$ phase to be $\delta=1.39\: \pi$~\cite{newosc}, and have assumed the Majorana phases in the PMNS matrix to be zero.  A similar analysis can be performed for an inverted mass hierarchy with the corresponding best fit oscillation parameters. 

\begin{figure}[t!]
\centering
\includegraphics[width=8.5cm]{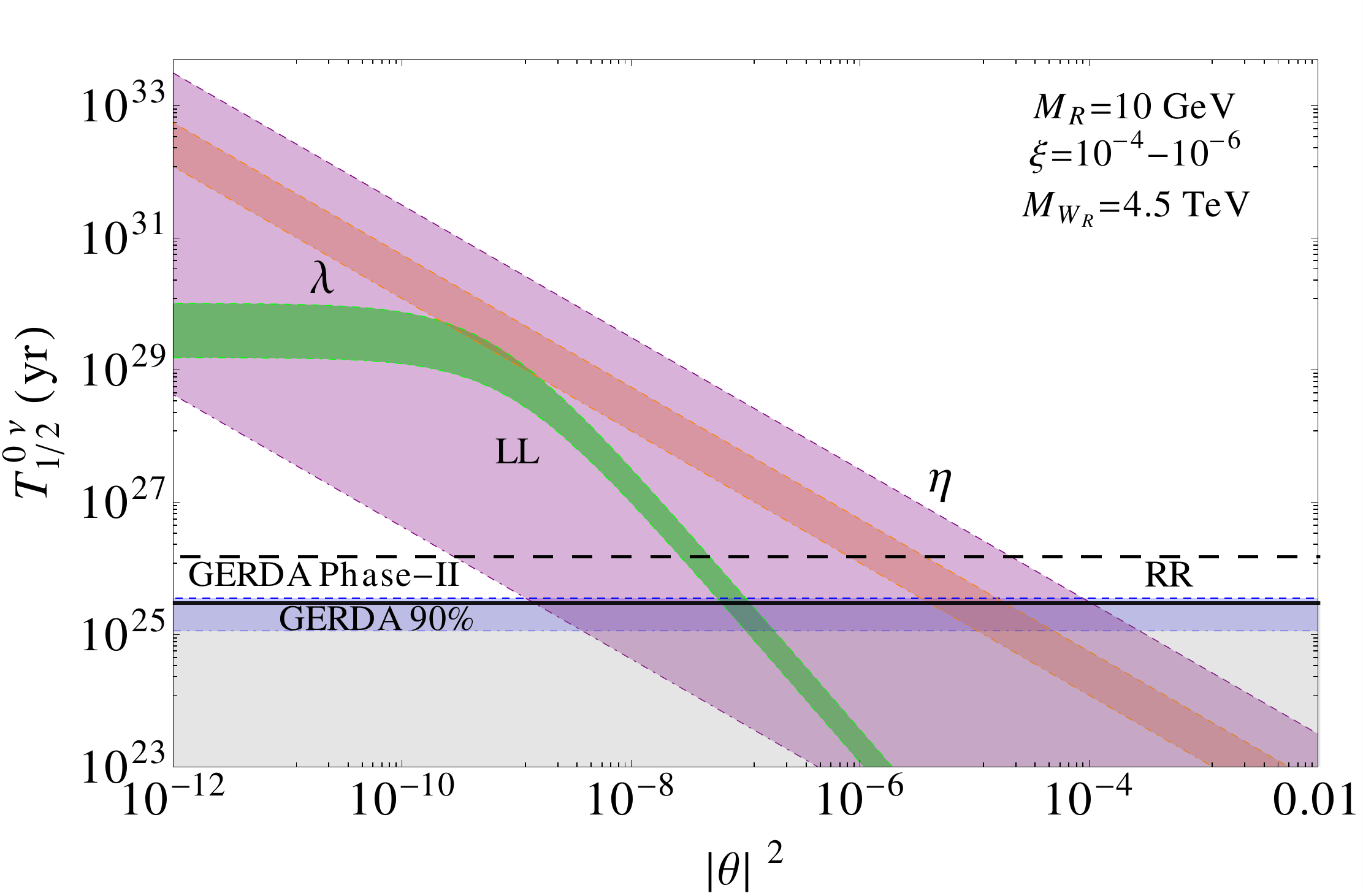}
\includegraphics[width=8.5cm]{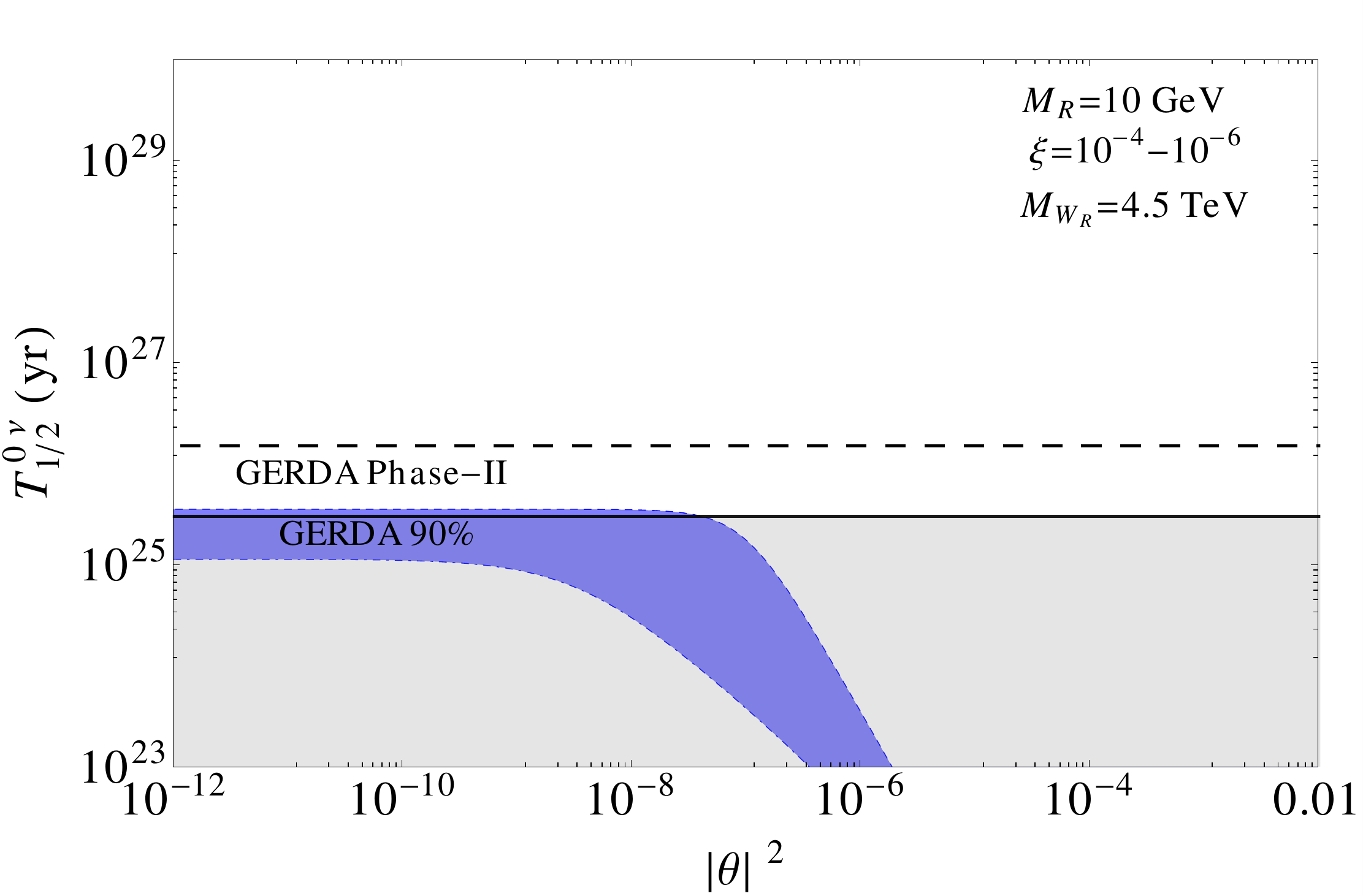}
\\
\includegraphics[width=8.5cm]{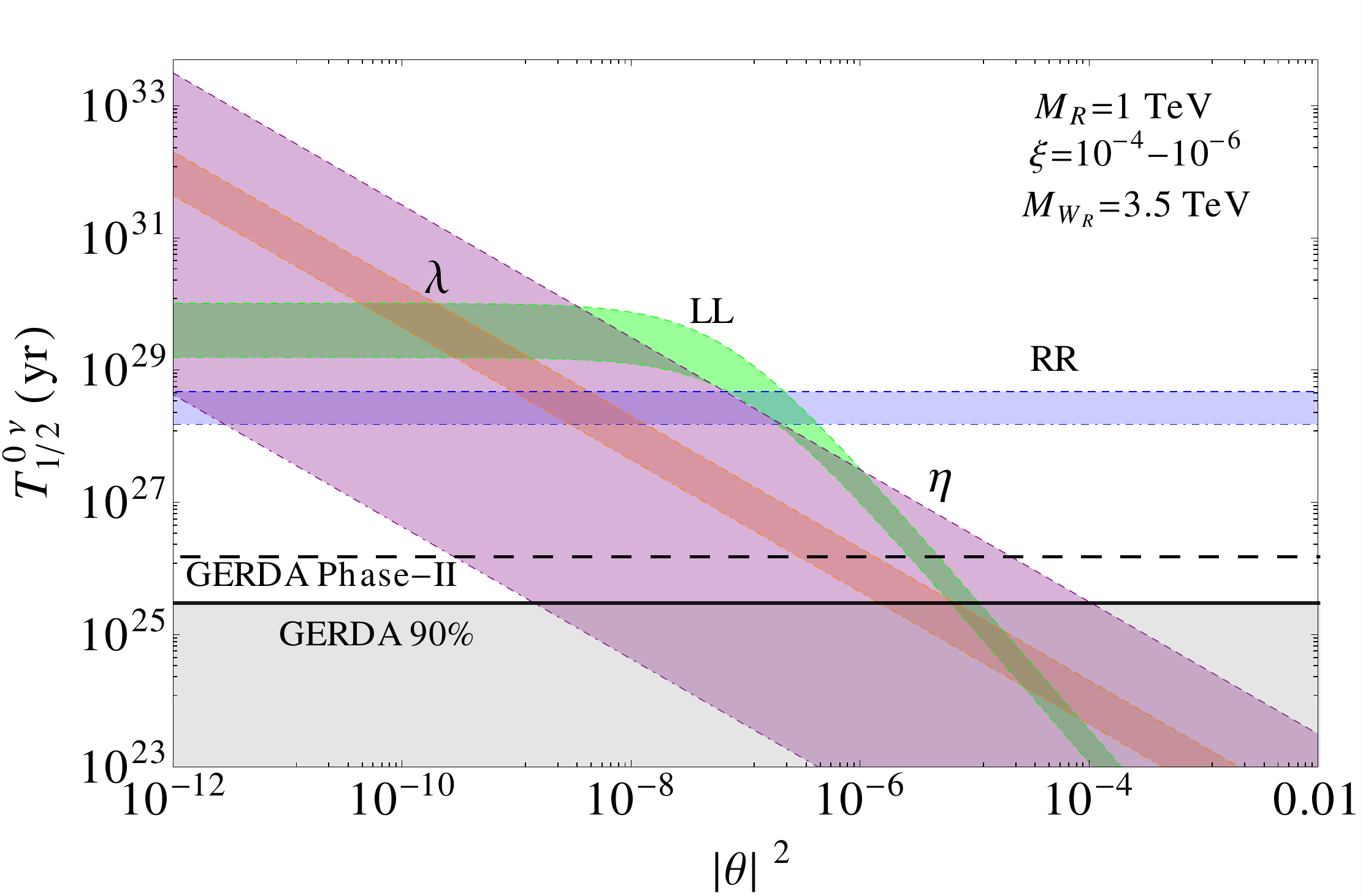}
\includegraphics[width=8.5cm]{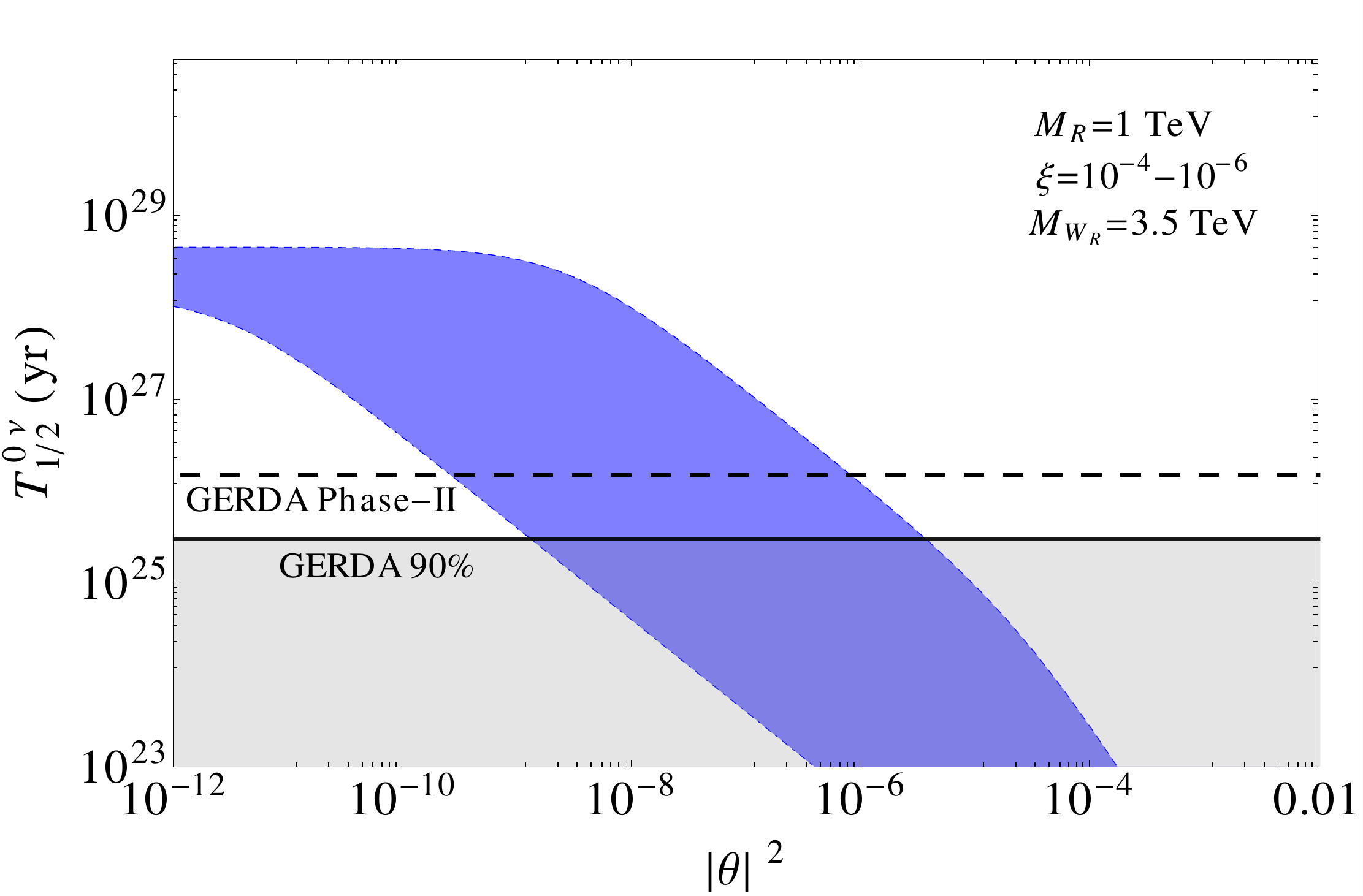}
\caption{The left panels show the individual contributions discussed in Section~\ref{diffcont} 
to the $0\nu\beta\beta$ half life of $^{76}$Ge as a function of the light-heavy neutrino mixing 
parameter. The right panels show the total contribution, as obtained from Eq.~(\ref{half}). Here we have considered two benchmark scenarios: (i) $M_R=10$ GeV and $M_{W_R}=4.5$ TeV (top panel), and (ii) $M_R=1$ TeV and $M_{W_R}=3.5$ TeV (bottom panel). In both cases, the LH-RH gauge boson mixing parameter $\xi$ is varied between $10^{-4}$--$10^{-6}$. The gray shaded region is excluded by the GERDA-I results~\cite{gerda}, while the dashed horizontal line shows the GERDA-II projected sensitivity~\cite{gerda2}.}
\label{fig2}
\end{figure}

Within the simplified framework described above, we show in Figure~\ref{fig2} the different  contributions to the $0\nu \beta \beta$ half-life (left panel) as well as the total  contribution (right panel) as a function of the light-heavy 
mixing parameter $\theta$. Note that in the canonical seesaw limit, the order parameter $\theta$ is severely constrained by the smallness of light neutrino mass: $\theta^2\lsim 10^{-10}(1~{\rm GeV}/M_R)$ for a light neutrino mass of 0.1 eV. Hence, all the RH neutrino contributions that depend on $\theta$ become negligibly small. However, in the presence of cancellations in Eq.~\ref{eq1}, the stringent bound on $\theta$ can be circumvented, leading to the possibility of large $\lambda$, $\eta$ and RH neutrino contribution via $W_L-W_L$ channel [cf. Figure~\ref{fig1}(c)]. Henceforth, 
we have assumed this to be the case, unless otherwise specified.   

In order to demonstrate the interplay between the various contributions 
and to see whether an individual contribution by itself can saturate the current experimental bound, we have considered two benchmark values of the 
RH neutrino mass: (i) $M_R$ = 10 GeV (top panel) and (ii) 1 TeV (bottom panel), and compare the half-life predictions for $^{76}$Ge with the 90\% CL combined limit from GERDA~\cite{gerda}. A similar analysis can be performed 
for the $^{136}$Xe isotope. The different bands in Figure~\ref{fig2} (left panels) correspond to 
the individual contributions in Eq.~(\ref{half}), including the relevant NME uncertainties. 
 The band denoted as LL (green) includes the LH contributions from the light and heavy neutrinos via $W_L-W_L$ channel. Here we have assumed a hierarchical light neutrino spectrum and have just used the oscillation data to derive the canonical light neutrino contribution [cf. Eq.~(\ref{etanu})], assuming  that it is independent of the mixing parameter $\theta$, 
which is possible in the presence of 
cancellations, as mentioned earlier. An  explicit model with the full flavor structure of $M_D$ and $M_R$ will be given in Section.~\ref{largemix}. Thus, the $\theta$-dependence of the LL contribution comes solely due to the diagram shown in Figure~\ref{fig1}(c), which becomes dominant over the canonical contribution shown in Figure~\ref{fig1}(a) for large values of $\theta$.  The purely RH contribution [cf. Figure~\ref{fig1}(b)] is always independent of the active-sterile mixing, as shown by the horizontal RR (blue) band.  
The $\lambda$ (pink) and $\eta$ (purple) bands always depend on the mixing parameter $\theta$, but independent of the heavy neutrino mass $M_R$. For the $\eta$ contribution, we have also 
varied the parameter $\xi$ between $10^{-4}$--$10^{-6}$. 

For case (i) with $M_R=10$ GeV (top panels), the RR contribution is the dominant one for small values of mixing, and violates the current GERDA bound~\cite{gerda} (solid horizontal line) for $M_{W_R} < 4.3$ TeV. Hence, we have considered $M_{W_R}=4.5$ TeV for this case. The RR-dominance in the small-$\theta$ region is also reflected in the total contribution (top, right panel), whereas  for higher $\theta$ values, the $\eta$ contribution dominates and saturates the current GERDA limit. For case (ii) with $M_R=1$ TeV and $M_{W_R}=3.5$ TeV, the RR contribution is much smaller, and cannot saturate the current bound, or even the future projected bound from GERDA-II~\cite{gerda2} (horizontal dashed line). The LL contribution can saturate the current limit for $|\theta|^2 \sim (6.1 \times 10^{-8}-1.1 \times 10^{-7})$ in case (i) and $|\theta|^2 \sim (6.1 \times 10^{-6}-1.1 \times 10^{-5})$  in case (ii). In case (ii), the $\lambda$ contribution becomes dominant over the RR and LL  contribution for $|\theta|^2> 10^{-8}$ and $|\theta|^2> 10^{-9}$  respectively,  and saturates the current bound at $|\theta|^2 \sim ( 1.46-6.75) \times 10^{-6}$.  However, in both cases (i) and (ii), the $\eta$ contribution could become dominant over all other contributions for a relatively larger value of $\xi \sim  10^{-4}$, and could saturate the GERDA limit for a much smaller value of $|\theta|^2 \sim 1.42 \times 10^{-9}$. Thus, including the $\eta$ contribution leads to a much stronger upper limit on the mixing parameter $|\theta|^2$, as can be seen from Figure~\ref{fig2} (right panels), where we have shown the total contribution, as given by Eq.~(\ref{half}). The improved upper limit on the active-sterile neutrino mixing is discussed further in the following section.

\subsection{Improved Limit on Light-Heavy Neutrino Mixing} \label{sec:IIIA}
\begin{figure}[t!]
\centering
\includegraphics[width=8cm]{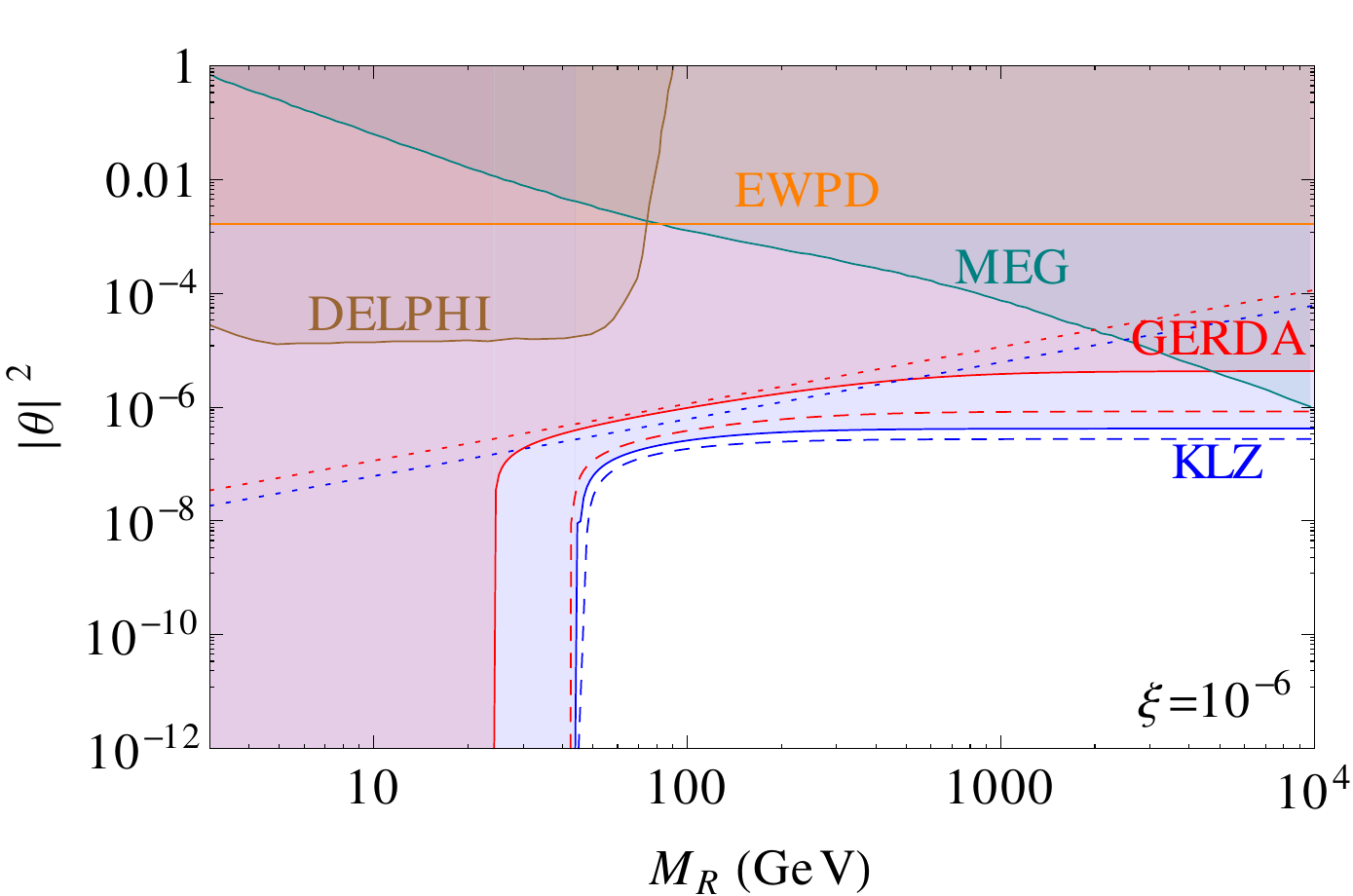}
\includegraphics[width=8cm]{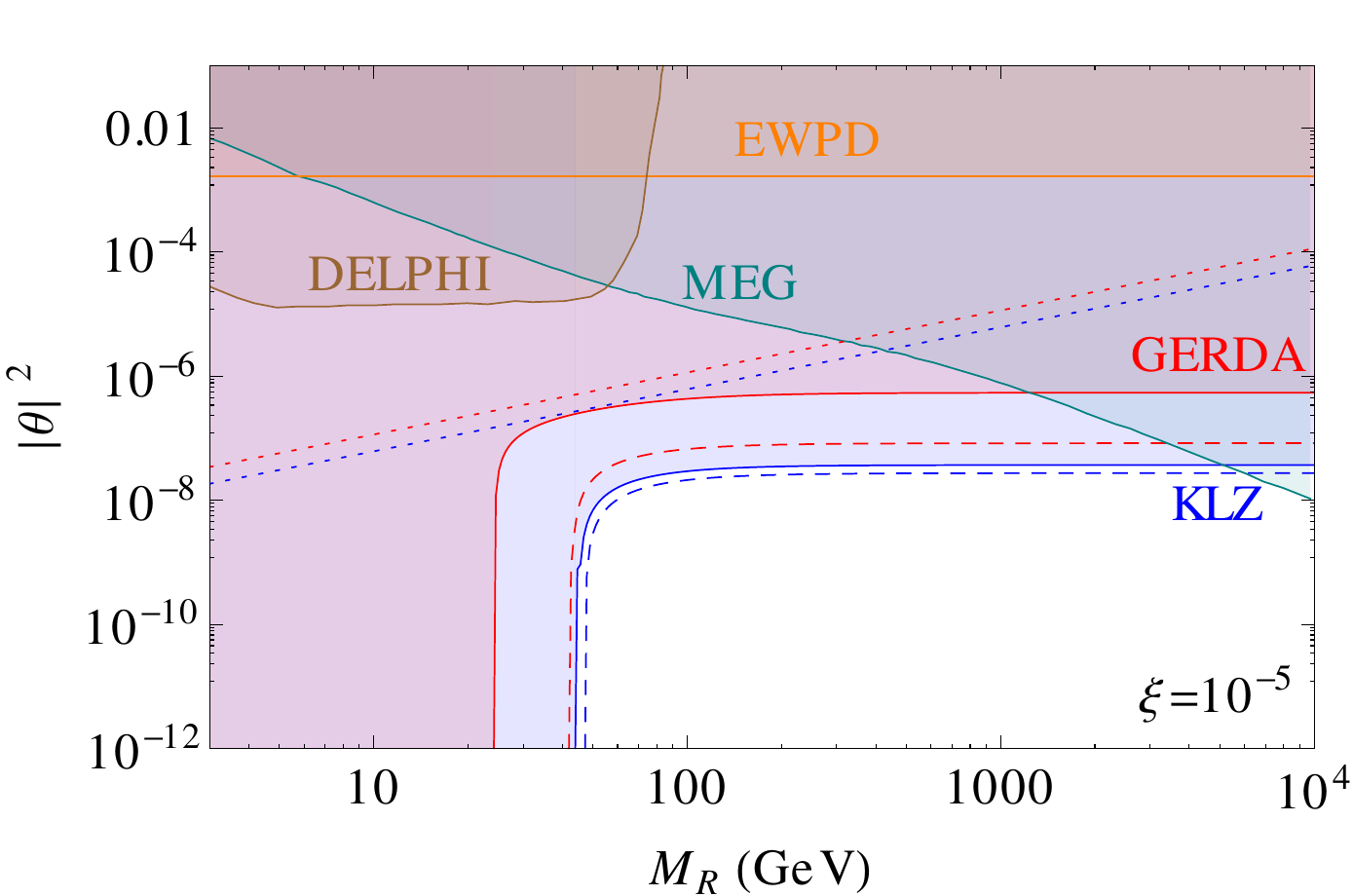}\\
\includegraphics[width=8cm]{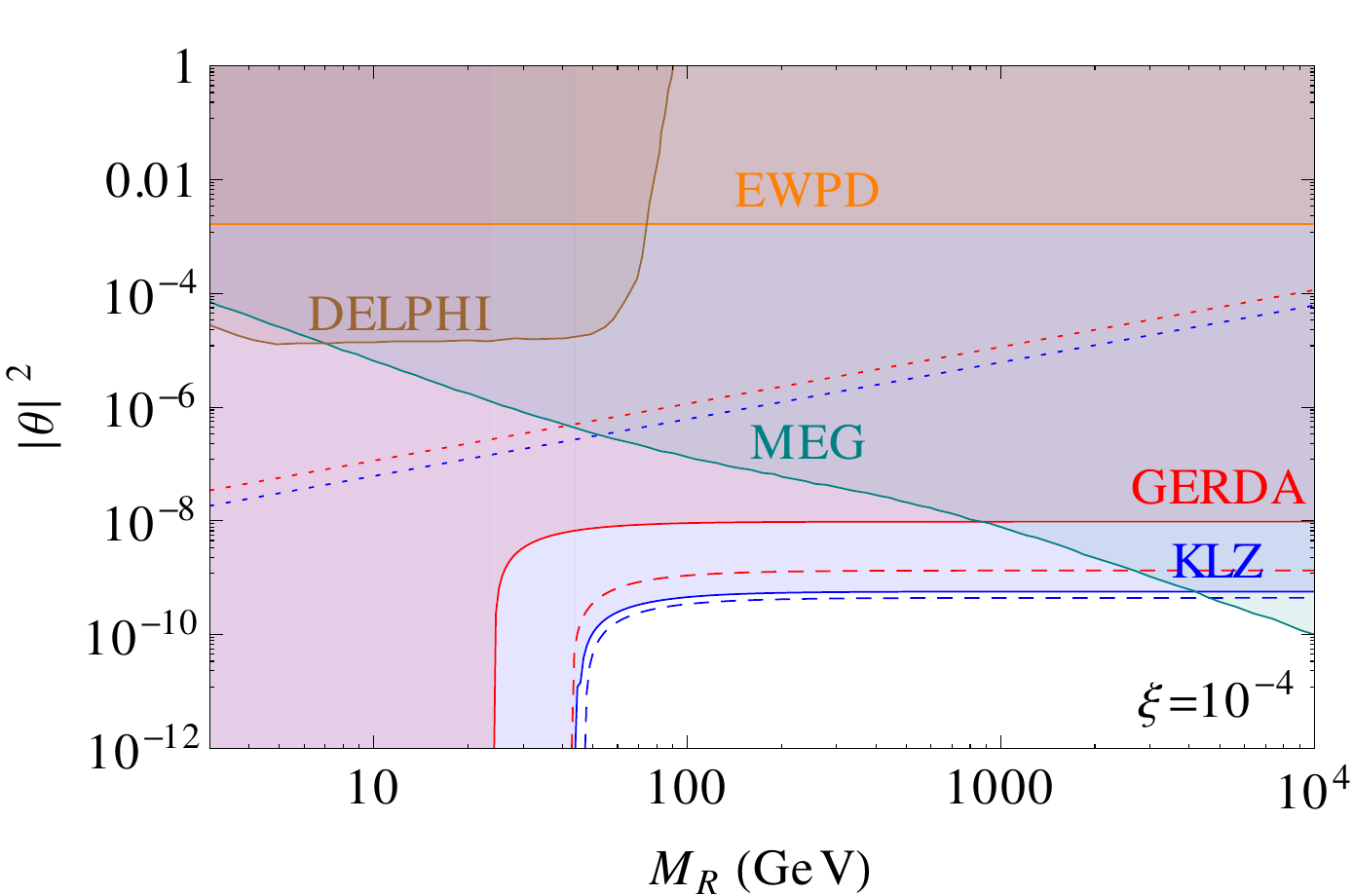}
\caption{The allowed values of the light-heavy neutrino mixing (white region) as a function of the heavy neutrino mass in the minimal LRSM for $M_{W_R}=3.5$ TeV and for different values of the gauge boson mixing parameter $\xi$. The dotted lines show the corresponding limits in the SM seesaw, i.e. derived from the LL contribution {\it only}. The shaded regions show the various exclusion limits; for details, see text. }
\label{fig3}
\end{figure}

In the simplified framework considered in this section, the relevant mixing matrices 
in the $\lambda$ and $\eta$ amplitudes given by Eqs. (\ref{lam}) and (\ref{eta}) respectively 
can be expressed  in terms of $T \simeq  -\theta^* U_{\nu}$, and $S \simeq \theta \, V_R = \theta U_\nu$. 
Thus, for a given value of the $W_R$ mass and the $W_L$--$W_R$ mixing parameter $\xi$, we can derive constraints in the $(M_R,\theta)$ parameter space using 
the experimental lower limits on $T_{1/2}^{0\nu}$ as given in Eq.~(\ref{GeXe}). 
Our results are shown in Figure~\ref{fig3} for the exclusion regions derived from both GERDA (red shaded) and KamLAND-Zen (KLZ, blue shaded) data. Here the solid (dashed) lines 
show the conservative (optimistic) limits, including the uncertainties due to NMEs.  
For illustration, we have fixed $M_{W_R}=3.5$ TeV, and have considered three representative values for the gauge boson mixing parameter $\xi$: $10^{-6}$, $10^{-5}$, $10^{-4}$, all of which are well within its current experimental limit given in Section~\ref{rev}. We note that for small values of $M_R$, the RR contribution given by Eq.~(\ref{RR}) becomes dominant irrespective of the mixing, 
and saturates the current experimental bound on $0\nu\beta\beta$ half-life. This is shown by the vertical lines in Figure~\ref{fig3}, which set a lower limit on $M_R$ between (45.1--47.3) GeV from $^{136}$Xe data and between (24.2--42.9) GeV from $^{76}$Ge data for the chosen value of $M_{W_R}$ and independent of other model parameters. In the absence of L-R symmetry, the upper limit on the active-sterile mixing parameter is governed by the amplitude ${\cal A}^L_{N_R}$ [cf.~Eq.~(\ref{lnr})], and the resulting limit becomes weaker as we go to higher values of $M_R$: 
 \be
|\theta|^2 \ \lsim \ \frac{1}{\sqrt{T^{0\nu}_{1/2} G^{0\nu}_{01}}} \frac{M_R}{m_p\mathcal{M}_{N}^{0\nu}\sum_i U_{ei}^2 } \; ,
\label{theta1}
\ee
as shown by the red (blue) dotted lines, derived from the constraints on the $^{76}$Ge ($^{136}$Xe) half-life. Here we have not shown the NME uncertainties for brevity. 
In the presence of a low-scale L-R symmetry, the additional $\lambda$ and $\eta$ contributions given by Eqs.~(\ref{lam}) and (\ref{eta}) which also depend on the active-sterile mixing parameter, lead to a stronger constraint on $|\theta|^2$. Note that both $\lambda$ and $\eta$ amplitudes are independent of the RH neutrino mass $M_R$, and hence, for larger $M_R$ masses where the LL contribution (\ref{lnr}) diminishes, the combined limit derived from including the $\lambda$ and $\eta$ contributions will be independent of $M_R$, as shown in Figure~\ref{fig3}. For larger values of the LH-RH gauge boson mixing $\xi$, the $\eta$-contribution becomes dominant, as already shown in Figure~\ref{fig2}, and therefore, the limit on $|\theta|^2$ will be dominantly governed by this term:  
\be
|\theta|^2 \ \lsim \ \frac{1}{T^{0\nu}_{1/2} G^{0\nu}_{01}} \frac{1}{\tan^2 \xi |\mathcal{M}^2_{\eta}|
} \; .
\label{theta2}
\ee 
Thus, the upper limit on $|\theta|^2$ becomes more stringent for larger values of $\xi$, as can be seen from Figure~\ref{fig3}. The limits derived here are at least one order of magnitude stronger than those obtained in the SM seesaw case (dotted lines).  

There are complementary constraints on the $(M_R,\theta)$ parameter space coming from LFV observables which also receive additional contributions in the LRSM (for detailed studies, see e.g.~\cite{barry, Cirigliano:2004mv}). In Figure~\ref{fig3}, we show the constraint from the LFV process $\mu\to e\gamma$ (green shaded region) in our simplified scenario assuming that the mixings in the muon sector are same as those in the electron sector. These limits were derived by comparing the MEG limit on BR$(\mu\to e\gamma)<5.7\times 10^{-13}$ at 90\% CL~\cite{Adam:2013mnn} with the theoretical prediction
\be
{\rm BR}(\mu \to e\gamma) \ = \ \frac{\alpha_w^3 s_w^2}{256\pi^2}\frac{m_\mu^4}{M_{W_L}^4}\frac{m_\mu}{\Gamma_\mu}\left(|G_L^\gamma|^2+|G_R^\gamma|^2\right) \; , \label{mueg}
\ee
where $m_\mu$ and $\Gamma_\mu$ are respectively the mass and width of the muon, $s_w\equiv \sin\theta_w$ is the weak mixing parameter and $\alpha_w \equiv g^2/4\pi$ is the weak coupling strength. The form factors $G_{L,R}^\gamma$ are given by~\cite{barry} 
\be
G_L^\gamma & = & \sum_i\left[V_{\mu i} V^*_{ei}\left\{|\xi|^2G_1^\gamma(x_i)+\left(\frac{M_{W_L}}{M_{W_R}}\right)^2 G_1^\gamma(y_i)\right\}-S^*_{\mu i}V^*_{ei} \xi \frac{M_i}{m_\mu} G_2^\gamma(x_i)\right]\;, \label{mueg1}\\
G_R^\gamma &=& \sum_i\left[S^*_{\mu i}S_{ei} G_1^\gamma(x_i) - V_{\mu i}S_{ei} \xi \frac{M_i}{m_\mu} G_2^\gamma(x_i)\right] \; , \label{mueg2} 
\ee
where $x_i\equiv (M_i/M_{W_L})^2$, $y_i\equiv (M_i/M_{W_R})^2$, and the loop functions $G^\gamma_{1,2}(x)$ are defined as 
\be
G_1^\gamma(x) & = & -\frac{x(2x^2+5x-1)}{4(1-x)^3} - \frac{3x^3}{2(1-x)^4}\ln x \; ,\\
G_2^\gamma(x) & = & \frac{x^2-11x+4}{2(1-x)^2} - \frac{3x^2}{(1-x)^3}\ln x \; . \label{g2g}
\ee 
From the last two terms on the RHS of Eqs.~(\ref{mueg1}) and (\ref{mueg2}), we see that for relatively large values of the mixing parameters $\theta$ and $\xi$, the LFV rate BR($\mu\to e\gamma$) increases with $M_R$, and therefore, the LFV bound becomes stronger for larger $M_R$ values, as depicted in Figure~\ref{fig3}. In this sense, the $0\nu\beta\beta$ and LFV constraints in the large mixing regime of LRSM are truly complementary to each other.  However, it is interesting to observe that within the simplified framework considered here, the $\mu\to e\gamma$ rate depends on the combination $\sum_i U_{\mu i} U_{ei}$, as appearing in the last two terms on the RHS of Eqs.~(\ref{mueg1}) and (\ref{mueg2}), which vanishes for the Dirac $C\!P$ phase $\delta=n\pi$ (with $n=0,1,2,\cdots$), irrespective of the light neutrino mass hierarchy. In these cases, there is no $\mu\to e\gamma$ LFV constraint in Figure~\ref{fig3}. On the other hand, the $\lambda$ and $\eta$ contributions depend on the combination $\sum_i U_{ei}T^*_{ei}$ which is independent of the PMNS parameters in our case, and hence, the $0\nu\beta\beta$ constraints shown in Figure~\ref{fig3} are more robust. 

Apart from the LFV constraints, there exist other constraints on the $(M_R,\theta)$ parameter space from direct and indirect searches for heavy neutrinos. Some of these complementary constraints, namely the DELPHI limit on $Z$-decays~\cite{Abreu:1996pa} (brown shaded) and the indirect limit from EWPD~\cite{Blas:2013ana} (orange shaded) are also shown in Figure~\ref{fig3} for comparison with the $0\nu\beta\beta$ constraints derived here. The other existing limits from direct searches at LEP~\cite{Achard:2001qv} and at the LHC~\cite{LHC-LL} as well as from low-energy observables~\cite{atre} are all weaker than the limits shown here.

\subsection{Improved Limit on the Gauge Boson Mixing} \label{sec:IIIB}
\begin{figure}[t!]
\centering
\includegraphics[width=8.5cm]{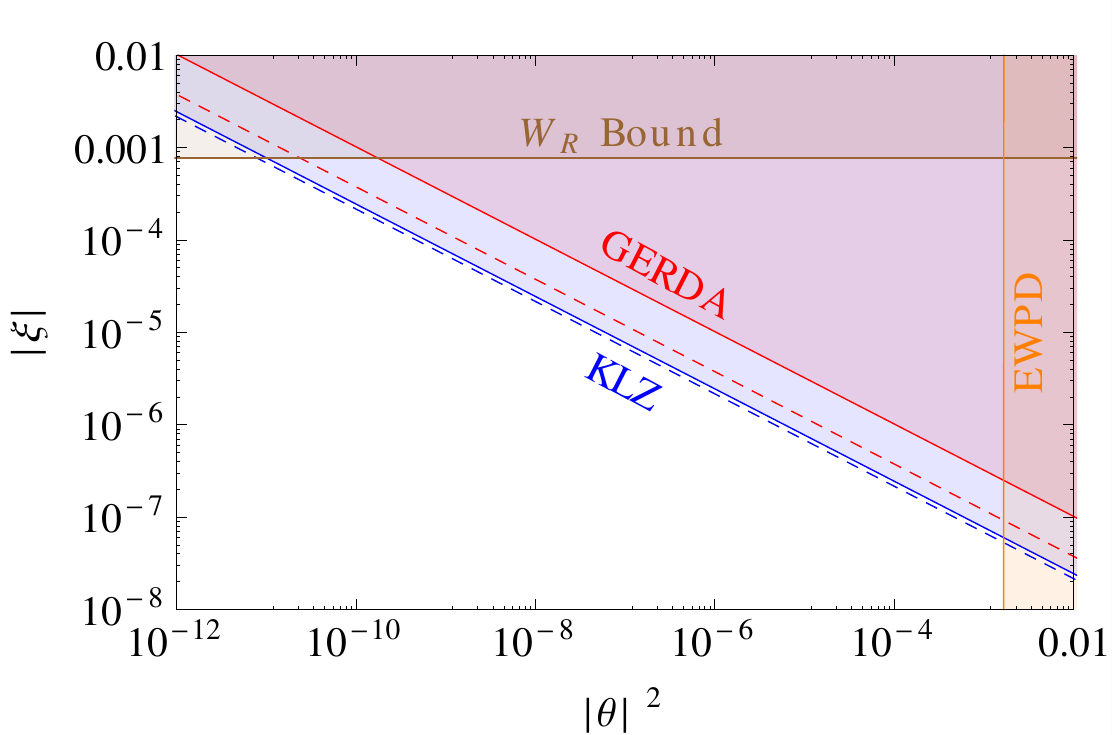}
\caption{Upper bound on the $W_L$-$W_R$ gauge boson mixing parameter 
$\xi$ from $0\nu\beta\beta$ constraints. The region between the solid and dashed slanted lines corresponds to the NME uncertainties. The horizontal line shows the current limit on $|\xi|$ from the lower limit on $M_{W_R}$, and the vertical line shows the indirect limit on the active-sterile mixing in the electron sector from EWPD.}
\label{fig4}
\end{figure}

In the limit when the $\eta$-contribution by itself saturates the current $0\nu\beta\beta$ bound, we can derive exclusion regions in the $(\theta,\xi)$ mixing plane using Eq.~(\ref{theta2}). 
This is shown in Figure~\ref{fig4} where the red (blue) shaded region is excluded from the 
$0\nu\beta\beta$ half-life limits from GERDA (KamLAND-Zen), independent of the other model parameters. The dashed lines correspond to the NME uncertainties. For comparison, the current limit on $|\xi|$ derived from the lower limit on $M_{W_R}>2.9$ TeV~\cite{Bertolini:2014sua} is also shown. The vertical line shows the indirect limit on the active-sterile mixing in the electron sector from EWPD~\cite{Blas:2013ana}. It is clear that the $\eta$-contribution to $0\nu\beta\beta$ provides a significantly improved limit on the mixing parameters in certain regions of the LRSM parameter space.

It should be noted here that a large value of $\xi$ might also lead to a significantly enhanced 
electric dipole moment (EDM) of charged leptons through the $W_L$--$W_R$ mixing diagram 
at one-loop level~\cite{pal}: 
\be 
d_l \ \simeq \ (4.2\times 10^{-21}e~{\rm cm}) \: {\rm Im}\left[\xi V G_2^\gamma(x_i) V^\dag \left(\frac{M_D}{1~{\rm GeV}}\right)\right]_{ll} \; , \label{edm}
\ee
where the function $G_2^\gamma(x)$ is defined in Eq.~(\ref{g2g}). In particular, the recent ACME upper limit on the electron EDM $d_e < 8.7\times 10^{-29}~e$ cm at 90\% CL~\cite{edm}, 
requires an extremely small value of $\xi$ for a TeV-scale LRSM, unless Im$[\xi M_{D_{ee}}]\ll 1 $. In the simplified analysis presented in this section, all the quantities in Eq.~(\ref{edm}) were assumed to be real, and therefore, relatively large values of $\xi$ as shown in  Figures~\ref{fig3} and \ref{fig4} are still consistent with the EDM constraints.  In Section~\ref{largemix}, we will consider a specific model, where the model predictions for the electron EDM turn out to be very close to the current upper limit for relatively large values of mixing.

\section{Analysis with Specific Flavor Structures \label{threerh}} 
In this section we extend the general analysis of the previous section to the case of three RH neutrinos by considering some specific flavor structures which satisfy the light neutrino oscillation data in the type-I seesaw dominance. We will consider two scenarios, depending on whether the active-sterile neutrino mixing parameters governing the contributions shown by Figure~\ref{fig1} (c)-(e) are small or large.  
\subsection{With Small Mixing \label{smallmix}}
In the canonical type-I seesaw limit, where the active-sterile mixing parameters $\theta \sim M_D M^{-1}_R$ are small, 
the dominant new contribution to $0\nu \beta \beta$ in the LRSM comes from purely RH currents as shown in Figure~\ref{fig1}(b). In this case, Eq.~(\ref{half}) simplifies to \cite{chakra} 
\be
\frac{1}{T_{1/2}^{0\nu}} \ \simeq \ G^{0 \nu}_{0\nu}\Bigg(\Big|{\cal M}_\nu^{0\nu}\eta_\nu|^2 + \Big|{\cal M}_N^{0\nu}\eta_{N_R}^R\Big|^2\Bigg) \ \equiv \ G^{0 \nu}_{0\nu}|{\cal M}_\nu^{0\nu}|^2\left|\frac{m_{ee}^{(\nu+N)}}{m_e}\right|^2 \; ,
\label{totalrh}
\ee
where $|m^{\nu+N}_{ee}|^2=|m^{\nu}_{ee}|^2+|m^N_{ee}|^2$ is the total effective mass, and 
$m^N_{ee}$ is the effective mass corresponding to the RH neutrino exchange [cf.~Eq.~(\ref{etaRR})]: 
\be
m^N_{ee} \ = \ \langle p^2 \rangle \left(\frac{M_{W_L}}{M_{W_R}}\right)^4 \sum_i  \frac{{V_{ei}^*}^2}{M_i}
\; ,
\label{mNee}
\ee
with $\langle p^2\rangle = -m_e m_p \mathcal{M}_N^{0\nu}/\mathcal{M}_{\nu}^{0\nu}$ denoting the virtuality of the exchanged neutrino.  Note that Eq.~(\ref{mNee}) is valid only in the heavy neutrino limit: $M_i^2 \gg |\langle p^2 \rangle | \sim (100~{\rm MeV})^2$ which is implicitly assumed here.  

In order to establish a simple relation between the active and sterile neutrino 
mass eigenvalues, we consider a specific case with $U=V$ in Eq.~(\ref{V}) and $U^{\sf T}h_DU={\bf 1}$ for the Yukawa couplings in Eq.~(\ref{eq:mnu}). In this simplified case, diagonalizing both sides of Eq.~(\ref{eq:mnu}), we obtain 
\be
\widehat{M}_\nu \ = \ U^{\sf T} M_\nu U \ = \  -\frac{\kappa^2}{\sqrt 2}U^\dag M_R^{-1} U^* \ = -\frac{\kappa^2}{\sqrt 2}\widehat{M}_R^{-1} \; ,
\ee 
or,~$m_i \propto 1/M_i$,~i.e. the light neutrino mass eigenvalues are inversely proportional to the heavy neutrino masses~\cite{chakra}. 
Thus, for a normal hierarchy (NH) with $m_1$ as the smallest, we have $M_1$ as the largest, and the other two heavy neutrino masses can be expressed in terms of $M_1$ as $M_2/M_1=m_1/m_2$ and 
$M_3/M_1=m_1/m_3$. So the effective mass for the heavy neutrinos given by Eq.~(\ref{mNee}) can be rewritten as
\begin{eqnarray}
m_{ee}^N|_{\rm NH} \ = \ \frac{C_N}{M_1}\left(c_{12}^2c_{13}^2+\frac{m_2}{m_1}s_{12}^2c_{13}^2e^{i\alpha'_1}+\frac{m_3}{m_1}s_{13}^2e^{i\alpha'_2}\right) \; ,
\label{nh}
\end{eqnarray} 
where $C_N = \langle p^2 \rangle (M_{W_L}/M_{W_R})^4$, $c_{ij}\equiv \cos\theta_{ij}$, $s_{ij}\equiv \sin\theta_{ij}$ and $\alpha'_{1,2}$ are two phases, related to the Dirac and Majorana 
$C\!P$ phases of the PMNS mixing matrix.  
On the other hand, for inverted hierarchy (IH) with $m_3$ as the smallest, $M_3$ will be the largest, and the other two right-handed neutrino masses can be expressed in therms of $M_3$ as $M_2/M_3=m_3/m_2$, and $M_1/M_3=m_3/m_1$. The effective mass for the heavy neutrinos given by Eq.~(\ref{mNee}) in this case will be 
\begin{eqnarray}
m_{ee}^N|_{\rm IH} \ = \ \frac{C_N}{M_3}\left(\frac{m_1}{m_3}c_{12}^2c_{13}^2+\frac{m_2}{m_3}s_{12}^2c_{13}^2e^{i\alpha'_1}+s_{13}^2e^{i\alpha'_2}\right) \; .
\label{ih}
\end{eqnarray} 

\begin{figure}[t!]
\includegraphics[width=8cm]{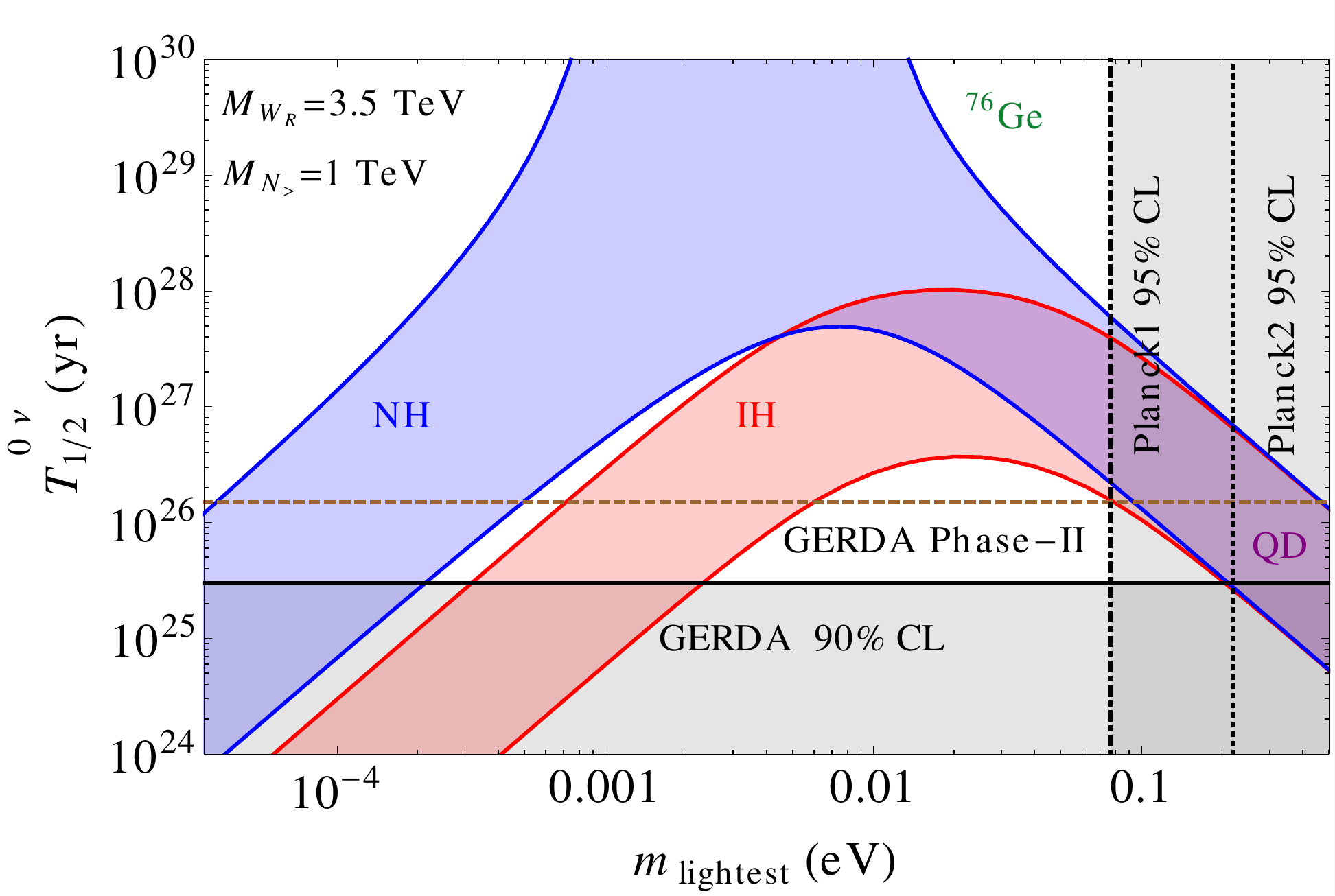}
\includegraphics[width=8cm]{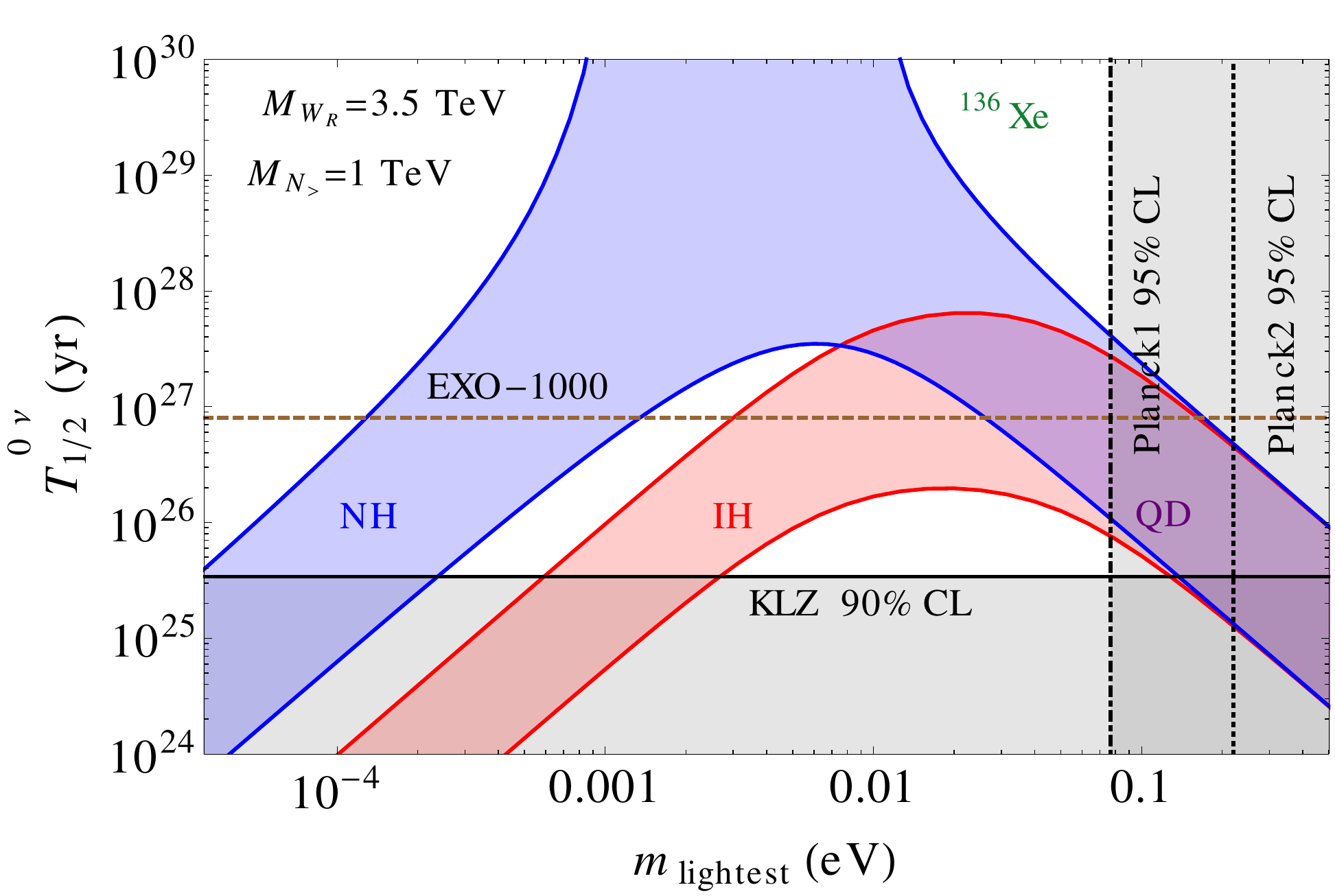}
\caption{
Variation of the half-life with the lightest neutrino mass in the type-I seesaw dominance with normal (NH), inverted (IH) and quasi-degenerate (QD) mass spectra for $^{76}$Ge (left panel) and $^{136}$Xe (right panel) isotopes. The vertical shaded regions are excluded at 95\% CL by different sets of Planck data. The horizontal shaded region is excluded at 90\% CL by GERDA-I~\cite{gerda} and 
KLZ~\cite{kamland} results. The horizontal dashed lines show the GERDA Phase-II~\cite{gerda2} and EXO-1000~\cite{exo1000} projected sensitivities. 
}  
\label{fig5}
\end{figure}

Using Eqs.~(\ref{nh}) and (\ref{ih}) in Eq.~(\ref{totalrh}), the predictions
for the half life for both NH and IH cases are shown in Figure~\ref{fig5}. 
Here we have chosen $M_{W_R}=3.5$ TeV and the heaviest neutrino mass
$M_{N_>}=1$ TeV for illustration. 
Note that the validity of the approximation $M_i^2 \gg |\langle p^2 \rangle|$ for the heavy neutrinos puts a restriction on how low $m_{\rm lightest}$   
can be. The scale of $m_{\rm lightest}$ used in Figure \ref{fig5} 
is in accordance with the above.   
We have considered the  variation of the three-neutrino oscillation 
parameters in their $3 \sigma$ range of a recent global fit~\cite{newosc},
and the Majorana phases are varied between 0 and $\pi$. 
We also include the updated NME uncertainties~\cite{petcov}, which gives the 
variation of $\langle p^2\rangle = -$(153--184 MeV)$^2$ for $^{76}$Ge and
$-$(157--185 MeV)$^2$ for $^{136}$Xe.  
Together, these variations result in the blue (red) shaded regions for the NH (IH) case. These predictions are to be compared with the current experimental lower limits Eq.~(\ref{GeXe}) 
shown by the solid horizontal lines, and the projected sensitivities~\cite{gerda2, exo1000} shown by the dashed horizontal lines. Also shown are the current upper limit on the lightest neutrino mass, as derived from the 95\% CL bounds on the sum of light neutrino masses in the quasi-degenerate (QD) regime~\cite{planck}: $\sum_i m_i < 0.23$~eV (Planck1) from the Planck+WMAP low-multipole polarization+high resolution CMB+baryon acoustic oscillation (BAO) data and assuming a standard $\Lambda$CDM model of cosmology, whereas the dashed vertical line shows
the limit without the BAO data set: $\sum_i m_i < 0.66$ eV (Planck2). 
Two important points can be inferred from Figure~\ref{fig5}: (i) The quasi-degenerate region in which the current experimental limit can be saturated by the light neutrino contribution alone is almost ruled out from the Planck data, thereby requiring an additional contribution in case a positive $0\nu\beta\beta$ signal is detected in near future. (ii) Including the purely RH contribution within the simplified framework adopted here, we obtain an absolute {\it lower} limit on the lightest neutrino mass, similar to that obtained in the type-II seesaw dominance~\cite{DGMR}. The exact value of this lower bound depends on the model parameters $M_{W_R}$ and $M_{N_>}$, and for the parameters chosen here, { we get $m_{\rm{lightest}} \gsim  (0.03-0.2)$ meV for $^{76}$Ge, $(0.05-0.2)$ for $^{136}$Xe for the IH case.} For the NH case, we cannot derive a lower limit since the NME uncertainties push the $m_{\rm lightest}$ values below the validity range of the analysis. 

\subsection{Large mixing \label{largemix}}
In this section, we study a specific case of TeV-scale LRSM scenario with large light-heavy neutrino mixing, thus extending our general analysis of Section.~\ref{gen} to  the three 
generation case, to explicitly demonstrate the importance of the $\lambda$ and $\eta$ diagrams. We consider the following Dirac and Majorana mass matrices:
\begin{equation}
M_D \ = \ \pmatrix {m_{e1} & m_{e2} & 0  \cr m_{\mu 1} &  m_{\mu 2} & 0 \cr m_{\tau 1} & m_{\tau 2} & 0 }\; , \qquad \qquad 
M_R \ = \ \pmatrix { 0 & M_{12} & 0 \cr M_{12}  & 0 & 0 \cr 0 & 0 &  M_{33}}\; .
\label{mtex}
\end{equation}
For this specific texture of $M_D$ and $M_R$,\footnote{Here we have assumed the discrete L-R symmetry and the $SU(2)_R$ gauge symmetry scales to be decoupled~\cite{chang}. Otherwise, the matrix structure of 
$M_D$ will be completely determined by the light neutrino mass matrix $M_\nu$ and the Majorana mass matrix $M_R$ by inverting the type-I seesaw formula [cf.~Eq.~(\ref{eq1})]~\cite{Nemevsek:2012iq}. } the different contributions to $0 \nu \beta \beta$, as listed in Eqs.~(\ref{etanu})-(\ref{etaeta}) become 
\be
\eta_{\nu} \ & = & \ -\frac{1}{m_e}(M_DM_R^{-1}M_D^{\sf T})_{ee} \ = \ -\frac{1}{m_e} \frac{2 \, m_{e1} m_{e2}}{M_{12}} \; , \label{seta1} \\
\eta^R_{N_R} \ & = & \ m_p \left(\frac{M_{W_L}}{M_{W_R}}\right)^4 (M^{-1}_R)_{ee} \ = \ 0\; , \label{seta2} \\
\eta^L_{N_R} \ &  = &\ m_p (M_D M^{-3}_R M^{\sf T}_D)_{ee} \ = \ m_p \frac{2 \, m_{e1}\,  m_{e2}}{M^3_{12}} \; , \label{seta3} \\
\eta_{\lambda} \ & = & \ -\left(\frac{M_{W_L}}{M_{W_R}}\right)^2 (M_D M^{-1}_R)_{ee} \ = \ 
-\left(\frac{M_{W_L}}{M_{W_R}}\right)^2 \frac{m_{e2}}{M_{12}} \;, \label{seta4} \\
\eta_{\eta} \ & = & \ -\tan \xi (M_D M^{-1}_R)_{ee} \ = \ - \tan \xi \, \frac{m_{e2}}{M_{12}} \; . \label{seta5}
\ee
Note that the purely RH current contribution is identically zero for the $M_R$ texture given by Eq.~(\ref{mtex}). Also, the $\eta$ and $\lambda$ contributions depend on the parameter 
$m_{e2}$, while the LL contributions $\eta_{\nu}$ and $\eta^L_{N_R}$ depend on the combination 
$m_{e1} \, m_{e2}$. Hence, a dominant $\eta$ and $\lambda$ contribution  
is possible to obtain in the limit $m_{e2}\gg m_{e1}$.

For the choice of $M_D$ and $M_R$ given in Eq.~\ref{mtex}, the light neutrino mass matrix in the type-I seesaw dominance has the following form: 
\be
M_{\nu} \ = \ -\frac{1}{M_{12}} \pmatrix{ 2 m_{e1}m_{e2} & m_{e2} m_{\mu 1}+m_{e1} m_{\mu 2} & m_{e2} m_{\tau 1}+m_{e1} m_{\tau 2} \cr
m_{e2} m_{\mu 1}+m_{e1} m_{\mu 2} & 2 m_{\mu 1}m_{\mu 2} &  m_{\mu 2} m_{\tau 1}+m_{\mu 1}m_{\tau 2}  \cr
m_{e2} m_{\mu 1}+m_{e1} m_{\mu 2} & m_{\mu 2} m_{\tau 1}+m_{\mu 1}m_{\tau 2} & 2 m_{\tau 1}m_{\tau 2}}
\label{lightnu}
\ee
It is evident from the above structure that, all the light neutrino masses vanish in the limit when either $m_{e1}\to 0$ or $m_{e2}\to 0$. Thus, even in the presence of a hierarchy $m_{e 2}\gg m_{e 1}$, a light neutrino mass matrix consistent with the oscillation data can be obtained. To see this in a simple way, we recast the Dirac mass matrix in terms of  the Casas-Ibarra parametrization 
\cite{casas}: 
\begin{equation}
M_D \ = \ U_\nu \sqrt{m_i}\: R^{\sf T} \sqrt{M_i}\:  V^{\sf T}_R,
\label{casiba}
\end{equation}
The matrix $V_R$ diagonalizing the Majorana mass matrix $M_R$ in Eq.~(\ref{mtex}) with real   positive mass eigenvalues $M_i$ has the following form:  
\begin{equation}
V_R=\pmatrix {-\frac{i}{\sqrt{2}} & \frac{1}{\sqrt{2}} & 0 \cr 
\frac{i}{\sqrt{2}} & \frac{1}{\sqrt{2}} & 0 \cr
0 & 0 & 1 }\, .
\label{vr}
\end{equation}
As the third RH neutrino with mass $M_{33}\gg M_{12}$ is decoupled, this scenario is essentially same as the two degenerate RH neutrino case with mass $M_R$. 

\begin{figure}[t!]
\includegraphics[width=10cm]{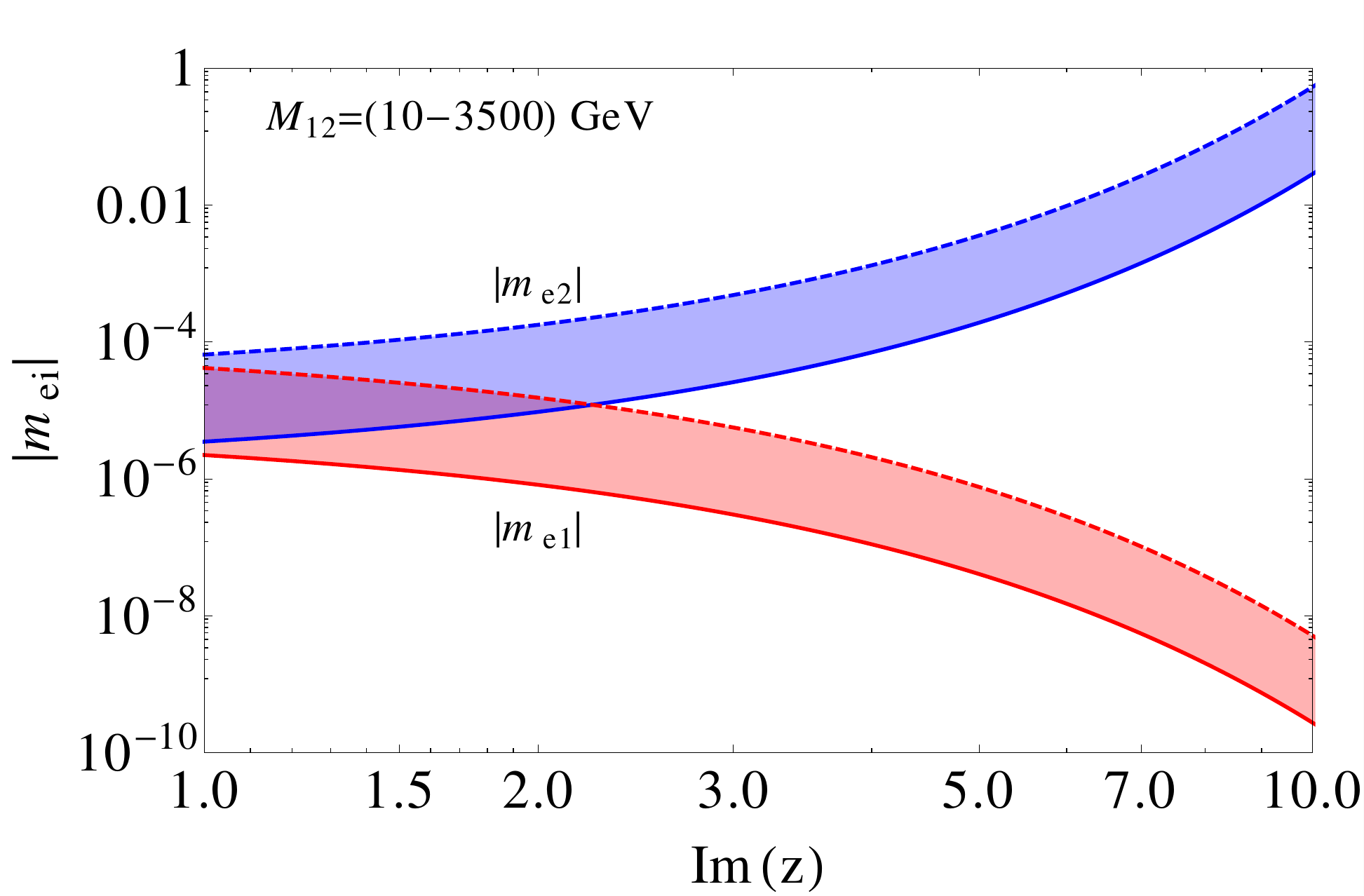}
\caption{Variation of the Dirac mass matrix elements $|m_{e1}|$ and $|m_{e2}|$ with Im($z$). 
The solid (dashed) lines are for $M_{12}=10$ GeV (3.5 TeV).   
}
\label{fig6}
\end{figure}

Let us take the orthogonal matrix $R$ in Eq.~(\ref{casiba}) to be of the following form: 
\begin{equation}
R=\pmatrix {0 & \cos z & -\sin z \cr 0 & \sin z & \cos z \cr 1 & 0 & 0 }\, ,
\label{R}
\end{equation}
where $z$ is a complex parameter. For illustrative purpose, we choose a NH mass spectrum with $m_1=0$ so that $m_2=\sqrt{\Delta m^2_{\rm sol}}$ and $m_3=\sqrt{\Delta m^2_{\rm atm}}$, where $\Delta m^2_{\rm sol}$ and $\Delta m^2_{\rm atm}$ are the solar and atmospheric neutrino mass-squared differences. The elements of the Dirac mass matrix  in Eq.~(\ref{casiba}) can now be written as   
\begin{eqnarray}
m_{l 1} \ & = & \ \sqrt{\frac{M_{12}}{2}} \Bigg[\left(U_{l2} \sqrt{m_2} \sin z + U_{l3} \sqrt{m_3} \cos z \right) -i\left (U_{l2} \sqrt{m_2} \cos z - U_{l3} \sqrt{m_3} \sin z \right) \Bigg]  \; ,\\
m_{l 2} \ &= & \ 
\sqrt{\frac{M_{12}}{2}} \Bigg[ \left(U_{l2} \sqrt{m_2} \sin z + U_{l3} \sqrt{m_3} \cos z \right)
+ i \left(U_{l2} \sqrt{m_2} \cos z - U_{l3} \sqrt{m_3} \sin z \right)\Bigg] \; ,
\label{mdi2}
\end{eqnarray}
with $l=e,\mu,\tau$. 
Using the best-fit values of the neutrino oscillation parameters from a recent global 
fit~\cite{newosc}, we show the variation of the two elements $|m_{e1}|$ and $|m_{e2}|$ as given by Eq.~(\ref{mdi2}) with the imaginary part of the parameter $z$. The results are shown in Figure~\ref{fig6}. Here we have taken Re($z$)=0.2 and have 
varied the mass parameter $M_{12}$ between 10 GeV and $M_{W_R}$ for illustration.  We find that increasing Im($z$) will result in an increase in $|m_{e2}|$ and a decrease in $|m_{e1}|$, as required to enhance the large mixing effects due to $\lambda$ and $\eta$ contributions to $0\nu\beta\beta$. Note that a small $m_{e1}$ is also desirable to have, especially for large $\xi$ values, in order to satisfy the stringent upper limit on electron EDM [cf. Eq.~(\ref{edm})].  
 
\begin{figure}[t!]
\includegraphics[width=0.49\textwidth, angle=0]{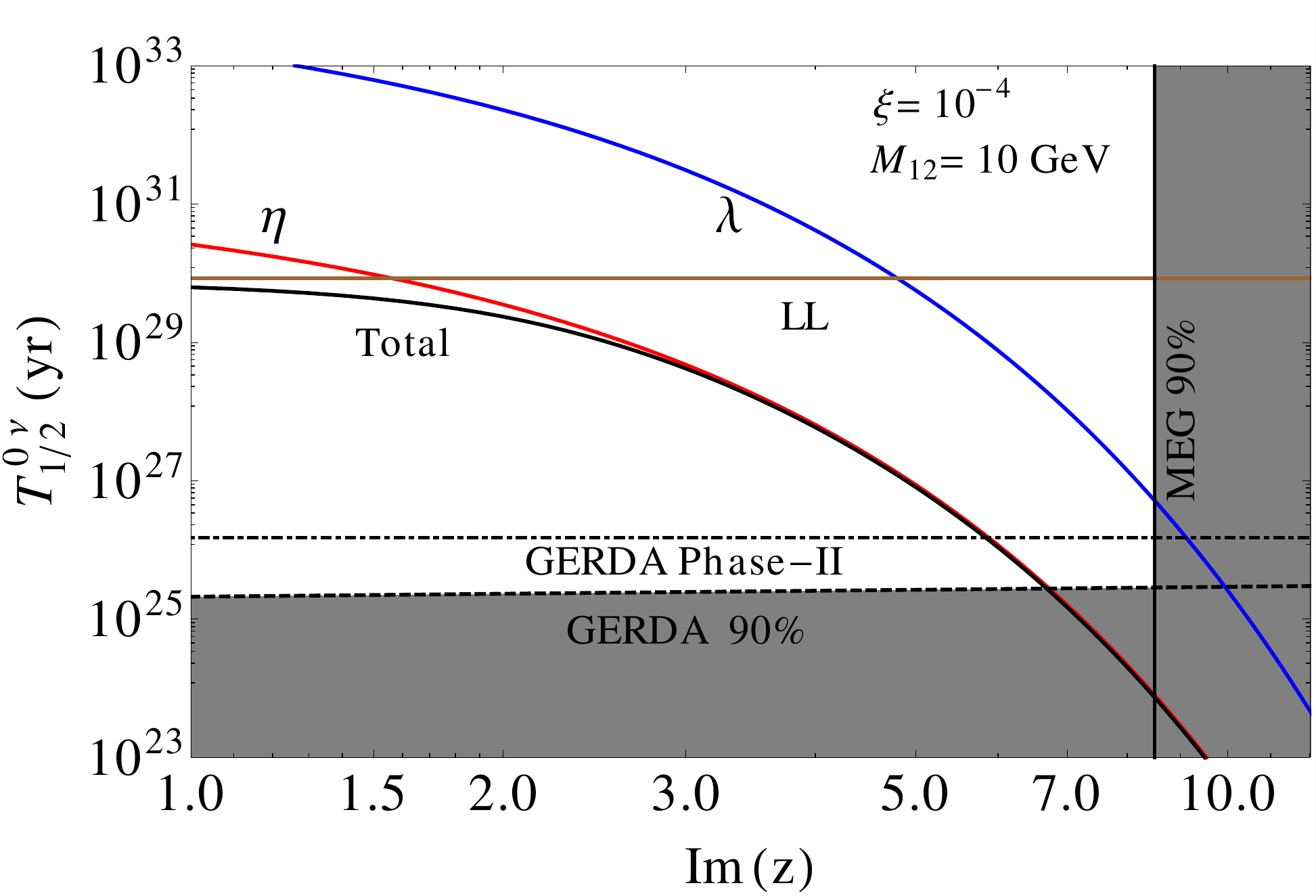}
\includegraphics[width=0.49\textwidth, angle=0]{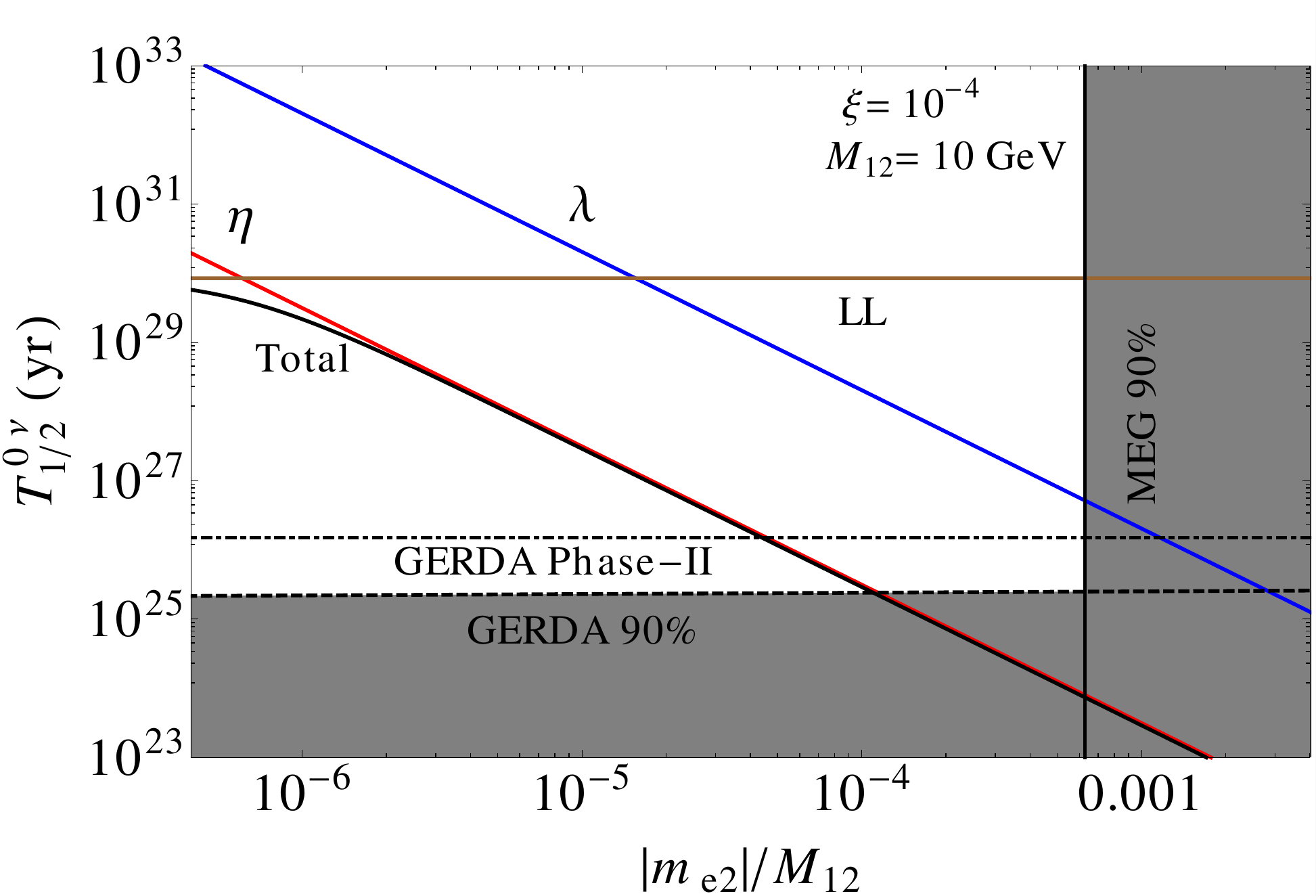}
\includegraphics[width=0.49\textwidth, angle=0]{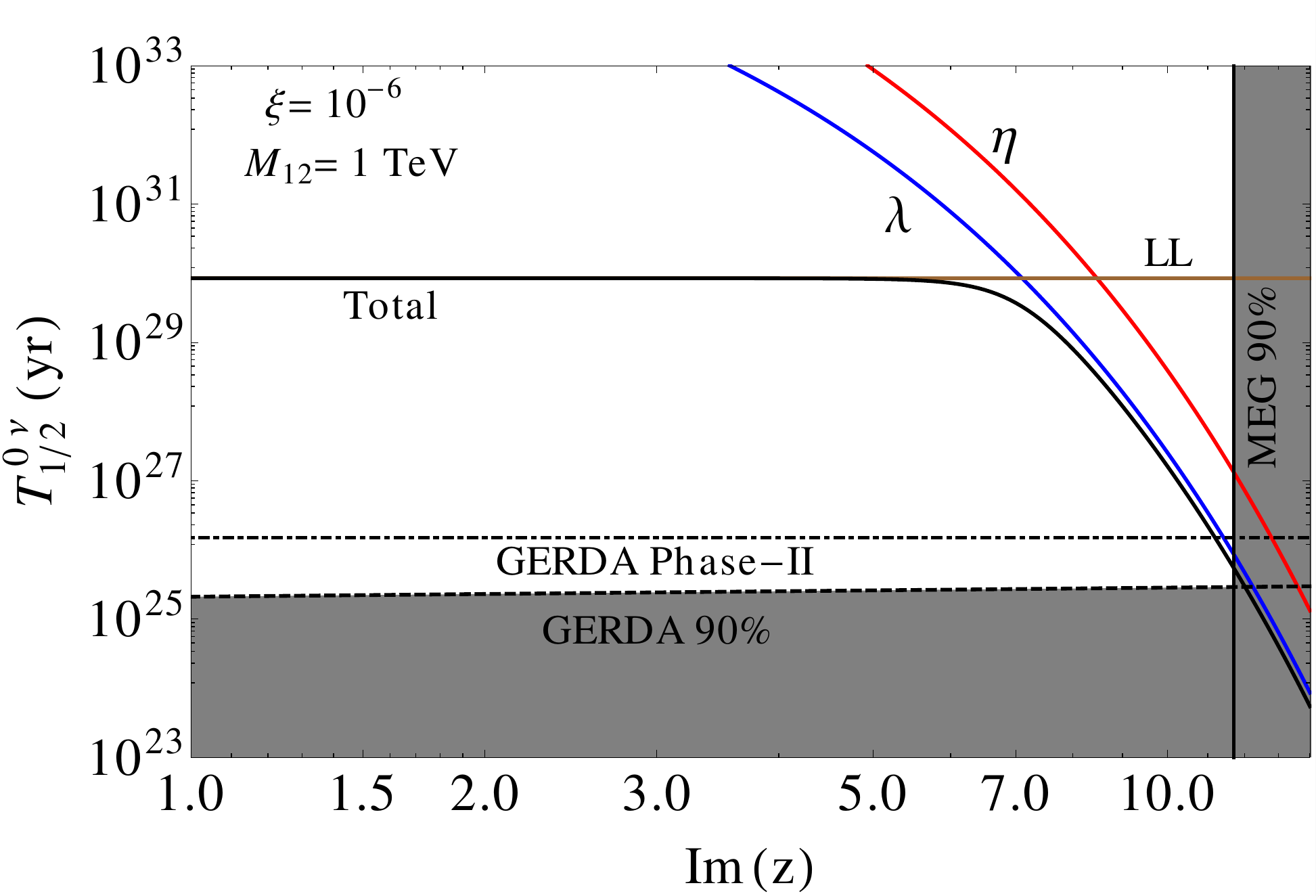}
\includegraphics[width=0.49\textwidth, angle=0]{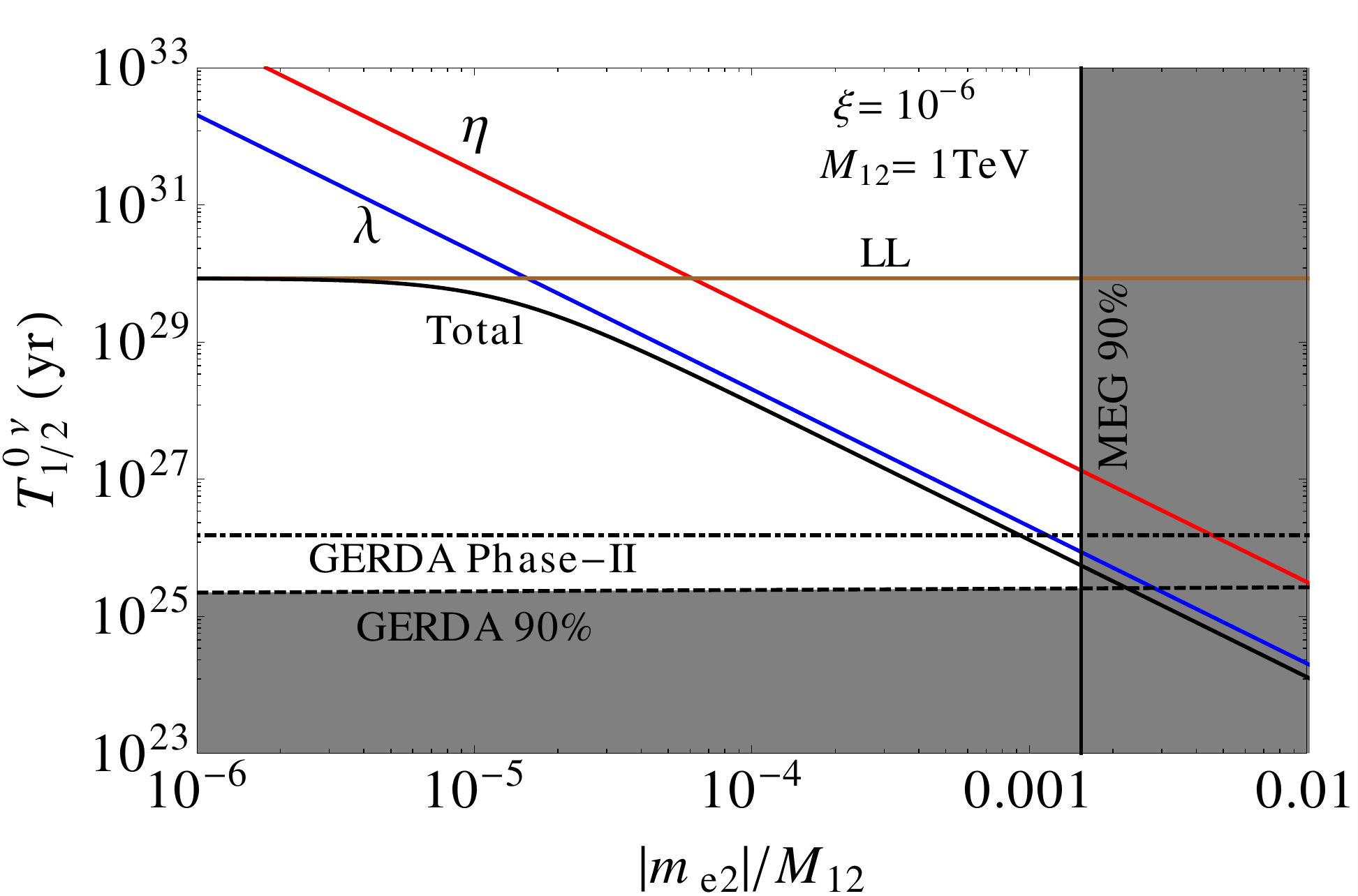}
\caption{Different contributions to $0\nu\beta\beta$ in our LRSM scenario with large mixing for two benchmark points (i) $\xi=10^{-4}$, $M_{12}=10$ GeV (top panels), and (ii) $\xi=10^{-6}$, $M_{12}=1$ TeV (bottom panels). In both cases, we have chosen $M_{W_R}=3.5$ TeV. The left panels show the variation of half-life as a function of the parameter Im($z$), whereas the right panels show the variation as a function of the light-heavy neutrino mixing parameter $m_{e2}/M_{12}$, analogous to that shown in Figure~\ref{fig2}.
{The black solid line corresponds to the total contribution}.  
}
\label{fig7}
\end{figure}

Assuming a large Im($z$) and the hierarchy $m_{l 2}\gg m_{l 1}$, the Dirac mass matrix given by Eq.~(\ref{mtex}) can be rewritten as  
\begin{equation}
M_D \ \simeq \  \pmatrix { \epsilon_1   & a_1 & 0 \cr \epsilon_2  & 
a_2  & 0 \cr \epsilon_3 & a_3 & 0 }
\label{mtex2}
\end{equation}
with $a_i \gg \epsilon_i$.  
In the limit $\epsilon_i \rightarrow 0$, the light neutrino masses identically vanish at tree-level, whereas the light-heavy neutrino mixing governed by $a_i$ can still be large. A natural embedding of this kind of texture in LRSM with an appropriate family symmetry is discussed in~\cite{DLM}, where the $\epsilon_i$'s can be treated as small perturbations from their symmetric limit $\epsilon_i\to 0$.  Observe that in the symmetric limit, all the $0\nu\beta\beta$ amplitudes vanish, except the $\lambda$ and $\eta$ terms [cf.~Eqs.~(\ref{seta4}) and (\ref{seta5})]. 
For our specific scenario, we show the different contributions to 
$0\nu \beta \beta$ in Fig.~\ref{fig7} as a function of Im($z$). 
Following the structure of the light neutrino mass matrix in Eq.~\ref{lightnu}, 
we  obtain the LL contribution to $0\nu \beta \beta$ around 
$T^{0\nu}_{1/2} \sim 10^{29}$ yr, independent of the value of Im($z$), as shown by the solid horizontal line in Figure~\ref{fig7}. This includes both $\eta_\nu$ and $\eta^L_{N_R}$ contributions, which depend on the product $m_{e1}m_{e2}= \epsilon_1 a_1$ [cf.~Eqs.~(\ref{seta1}) and (\ref{seta3})]. However, the $\eta^L_{N_R}$ term is extremely small in this case and the LL contribution is mostly governed 
by the light neutrino contribution. On the other hand, the $\eta$ and $\lambda$ contributions depend only on 
$a_1$ [cf.~Eqs.~(\ref{seta4}) and (\ref{seta5})], and hence, give the dominant contribution in this scenario, as shown in Figure~\ref{fig7}. 

The ratio of the half-lives corresponding to the $\eta$ and $\lambda$ contributions 
 is given by 
\begin{equation}
\frac{T^{0\nu}_{1/2}(\eta)}{T^{0\nu}_{1/2}(\lambda)} \ = \ \left|\frac{\mathcal{M}^{0\nu}_{\lambda}}{\mathcal{M}^{0\nu}_{\eta}}\right|^2 \left(\frac{M_{W_L}}{M_{W_R}}\right)^2 \frac{1}{\tan^2 \xi} \; .  
\end{equation}
Thus, for larger $\xi$ values, the $\eta$ contribution will be dominant, and can indeed saturate the present experimental limit. In Figure~\ref{fig7}, we show two benchmark points in the LRSM parameter space with (i) $\xi=10^{-4}$, $M_{12}=10$ GeV (upper panel) and (ii) $\xi=10^{-6}$ and $M_{12}=1$ TeV (lower panel), while $M_{W_R}=3.5$ TeV in both cases. We use the lower values of   NMEs given in Table.~\ref{tab1} and $\mathcal{M}_{\nu}=4.75$ . {For case (i), the $\eta$ contribution can saturate the present limit of $0\nu \beta \beta$ for Im($z$)=6.64.} This predicts an LFV rate BR$(\mu \to e\gamma)=1.24\times 10^{-14}$, which is still compatible with the current MEG limit, and could be probed with the upgraded MEG sensitivity~\cite{MEG2}. The predictions for other 
LFV rates, such as BR$(\tau \to e\gamma) \sim 10^{-22}$ and  BR$(\tau \to \mu \gamma) \sim  10^{-15}$, are extremely small to be observable in near future. For case (ii) with a higher mass $M_{12}= 1$ TeV, 
the LFV constraint is more stringent, and it is not possible to saturate the current $0\nu\beta\beta$ bound with the $\eta$ contribution.  However, in this case,  the $\lambda$ contribution can  be dominant and both the $\lambda$ and total contribution can reach the projected sensitivity of GERDA phase-II~\cite{gerda2}, while being marginally consistent with the current MEG limit.

Finally, we note that, due to the smallness of $m_{e1}$ in our model [cf.~Eq.~(\ref{mtex2})], the predictions for the electron EDM in case (i)  turn out to be $d_e \simeq 10^{-30}-10^{-33}e$~cm, depending on the exact value of Im($z$). These values of $d_e$ are still consistent with the upper limit from ACME: $d_e< 8.7\times 10^{-29}e$ cm~\cite{edm}. Similar results obtained for case (ii). Moreover, the exact prediction for $d_e$ will depend on the possible additional phases of $\xi$ and $V$ in Eq.~(\ref{edm}), without significantly affecting the $0\nu\beta\beta$ and LFV results, and hence, it is difficult to rule out this model solely based on the EDM constraints.

\section{ Conclusion \label{conclu}} 
We  have studied the predictions for the lepton number violating process of neutrinoless double beta decay ($0\nu \beta \beta$) within the framework of a TeV scale Left-Right symmetric theory assuming  type-I seesaw dominance.  
In this scenario, there exist several additional contributions to the $0\nu\beta\beta$ process, which depend on the light-heavy neutrino mixing and/or the $W_L-W_R$ gauge boson mixing. In the canonical type-I seesaw, the light-heavy neutrino mixing is severely constrained by the light neutrino mass constraint. In this case, the dominant additional contribution to $0\nu\beta\beta$ comes from the purely RH sector. However, the seesaw constraints on the mixing can be circumvented in presence of cancellations in the light neutrino mass matrix. This can be manifestly seen with specific textures of the Dirac and Majorana mass matrices, which could, in principle, be motivated by some symmetry. In this class of LRSM scenarios, the momentum-dependent contributions to $0\nu\beta\beta$ involving final state electrons with opposite helicities, i.e. the so-called  
$\lambda$ and $\eta$ contributions,  could be significant. This is the main result of this paper. 

To illustrate the large mixing effects on $0\nu\beta\beta$, we have first considered a simplified scenario (cf. Section~\ref{gen}) with a single heavy-neutrino mass scale and a common light-heavy neutrino mixing in the electron sector.
 We derive upper limits on this mixing parameter from the current experimental constraints on the $0\nu\beta\beta$ half-life. 
The main important point coming out of this analysis is that the $\eta$ contribution, which depends on the $W_L-W_R$ mixing $\xi$ and the light-heavy neutrino mixing $\theta$, could be dominant over other contributions in a wide range of the LRSM parameter space. This leads to stringent upper bounds on the mixing parameters $\theta$ and $\xi$, independent of the heavy neutrino mass. The improved upper limit on the light-heavy neutrino mixing is complementary to that obtained from LFV observables such as the $\mu\to e\gamma$ decay rate. 

Subsequently, we discuss a concrete TeV-scale LRSM with manifestly large mixing effects to demonstrate their importance for the $0\nu\beta\beta$ predictions. We show that for specific textures of the Dirac and Majorana mass matrices, the $\lambda$ and $\eta$ contributions could give the dominant contribution to $0\nu\beta\beta$ amplitude, while satisfying the light neutrino oscillation data. We consider two benchmark points to show the interplay between the $0\nu\beta\beta$ and LFV constraints, and find that in certain cases, the $\eta$ contribution could saturate the current experimental limit, while being consistent with LFV as well as EDM constraints. 

\section*{Acknowledgments}
We would like to thank Werner Rodejohann for helpful discussions. 
The work of P.S.B.D. is  supported by the Lancaster-Manchester-Sheffield Consortium for Fundamental 
 Physics under STFC grant ST/J000418/1. 
M.M. acknowledges partial support of the ITN INVISIBLES 
(Marie Curie Actions, PITN-A-2011-289442).


\end{document}